\documentclass[preprintnumbers,aps,prd,twocolumn,showpacs,showkeys,floatfix,preprintnumbers,letterpaper,amsmath,amssymb,superscriptaddress]{revtex4-1}
\usepackage{subfigure}
\usepackage{graphicx}
\usepackage{dcolumn}
\usepackage{epsfig}
\usepackage{amsmath, bm}
\usepackage{amsfonts}
\usepackage{color}
\usepackage{xcolor}
\usepackage{tabularx}
\usepackage{xspace}
\usepackage{float}
\usepackage[normalem]{ulem}

\newcommand{\jcap}{J. Cosmol. Astropart. Phys.}

\usepackage[T1]{fontenc}
\usepackage{ae,aecompl}
\newcommand{\be}{\begin{equation}}
\newcommand{\ee}{\end{equation}}
\newcommand{\bea}{\begin{eqnarray}}
\newcommand{\eea}{\end{eqnarray}}

\def\br{{\bf r}}
\def\bb{{\bf b}}
\def\bx{{\bf x}}

\def\bp{{\bf p}}
\def\bu{{\bf u}}

\def\bv{{\bf v}}
\def\bq{{\bf q}}
\def\bg{{\bf g}}

\def\bF{{\bf F}}

\def\G{{\cal G}}
\def\X{{\cal X}}
\def\IR{\mathbb{R}}

\def\rhobar{\bar{\rho}}

\def\trace{\mbox{trace }}

\def\Prob{\mbox{Prob }}
\def\PPhi{\Psi}
\def\D{\mathrm{D}}
\def\C{4\pi\G\rhobar} 
\def\MA{Monge-Ampère\xspace}
\def\MAG{Monge-Ampère gravity\xspace}

\begin{document}


\title{Monge Ampère gravity: \\ from the large deviation principle to cosmological simulations through optimal transport}

\author{Bruno Lévy}
\email{bruno.levy@inria.fr}
\affiliation{Inria Saclay Ile de France, Laboratoire de Mathématiques d'Orsay, 91405, Orsay, France }

\author{Yann Brenier}
\email{yann.brenier@ens.fr}
\affiliation{CNRS, Ecole Normale Supérieure, 45 rue d’Ulm, F-75230 PARIS cedex 05, France}

\author{Roya Mohayaee}
\email{mohayaee@iap.fr}
\affiliation{Sorbonne Université, CNRS, Institut d'Astrophysique de Paris, 98bis Bld Arago, 75014 Paris, France}
\affiliation{Rudolf Peierls Centre for Theoretical Physics, University of Oxford, Parks Road, Oxford OX1 3PU, United Kingdom}

\date{\today}

\begin{abstract}
 We study Monge-Ampère gravity (MAG) as an effective theory of cosmological structure formation through optimal transport theory. MAG is based on the Monge-Ampère equation, a nonlinear version of the Poisson equation, that relates the Hessian determinant of the potential to the density field. We explain how MAG emerges from a conditioned system of independent and indistinguishable Brownian particles, through the large deviation principle, in the continuum limit. To numerically explore this highly non-linear theory, we develop a novel N-body simulation method based on semi-discrete optimal transport. Our results obtained from the very first N-body simulation of Monge-Ampère gravity with over 100 millions particles show that on large scales, Monge-Ampère gravity is similar to the Newtonian gravity but favours the formation of anisotropic structures such as filaments. At small scales, MAG has a weaker clustering and is screened in high-density regions.  Although here we study the Monge-Ampère gravity as an effective rather than a fundamental theory, our novel highly-performant optimal transport algorithm can be used to run high-resolution simulations of a large class of modified theories of gravity, such as Galileons, in which the equations of motion are second-order and of Monge-Ampère type.
\end{abstract}


\pacs{}
\keywords{cosmology, optimal transport, Monge-Ampère, modified gravity, Galileons, extra dimensions}

\maketitle

\section{introduction}

\noindent
The elliptic Monge-Ampère (MA) equation:
\begin{equation}
    \det(\D^2\Phi(\br)) = \alpha \rho(\br)
    \label{eqn:MA1}
\end{equation}
is a nonlinear second-order partial differential equation, where the potential $\Phi$ and the density $\rho$ are functions of spatial coordinates $\br$,
$\D^2$ denotes the Hessian, and $\alpha$ is a constant scalar.
It has applications in optimal transport, meteorology, optics, oceanography and astrophysics. It exhibits some similarities with the Poisson equation,
\begin{equation}
    \Delta\Phi(\br)=\alpha\rho(\br),
    \label{eqn:Poisson1}
\end{equation}
that prescribes the sum of the eigenvalues of the Hessian $\Delta\Phi = \trace(D^2\Phi)$ whereas the \MA equation prescribes their product.

In one dimension, the two equations are the same. In higher dimension, the \MA equation
can be linearized around the identity to obtain the Poisson equation. The extra off-diagonal terms may be thought of as tidal tensors, yielding additional anisotropic components in the associated force. Unlike the Poisson equation, the \MA equation is invariant under affine symmetry which includes shear transformation. As we shall see, affine symmetry yields anisotropic structures -- {\it e.g.}\ ellipsoidal haloes, filaments and sheets -- more stable and abundant in Monge-Ampère gravity than in the Newtonian gravity.
Furthermore, unlike Newtonian gravity, MAG remains finite at short distances, which imply a screening mechanism in nonlinear regimes.
\\

The MA equation was first studied 200 years ago by Monge, and later generalised by Ampère. In the 90's, one of us, Brenier \cite{BrenierPFMR91} established the relation with another legacy of Monge: optimal mass transport \cite{Monge1784}, that corresponds to the gradient of the potential that solves the \MA equation.
The existence and regularity of solutions as well as ways of solving the \MA equation have been active areas of research \cite{figalliBook,opac-b1122739,OTON}.


In meteorology, replacing Poisson by \MA equation was a fruitful strategy:
Eliassen's semi-geostrophic model \cite{Eliassen48}, provides a successful framework to study atmospheric motion
with the formation of shocks and fronts \cite{Hoskins75,MC_GEO_1984}, and works better on large scales than the
older Poisson-based quasi-geostrophic model \cite{cullenBook}.

It has been shown by one of us, Yann Brenier \cite{brenier:hal-01137528} that Monge-Ampère gravity
can be derived from a system of Brownian particles through the application of large deviation principle (more on this later).

Monge-Ampère equation is also tightly related to optimal transport which has been shown to be a powerful \emph{inverse method} for reconstructing the initial condition of the Universe from the present data. The relation between the \MA equation, optimal transport and the least action principle \cite{DBLP:journals/nm/BenamouB00}
was used to design over the years increasingly efficient methods for early Universe reconstruction \cite{EURNature,EUR,levy_mnras_2021,vhauss_prl_2022,nikak_prl_2022,farnik-nikhil}, that may be thought of as variants of Peeble's action method \cite{1989ApJ...344L..53P,nusserbranchini}.  \\

In the present article, in contrast with the reconstruction methods mentioned above, we study for the first time the \emph{forward} dynamics governed by the Monge-Ampère equation, using high resolution numerical simulations. Here we only consider \MAG as an emergent rather than fundamental force, at cosmological scales, where convexity holds. The validity of \MAG at smaller scales (sub Mpcs), where convexity breaks down due to multi-streaming \citep{EUR}, shall be discussed in forthcoming works \cite{bonnefous}. \\

In the rest of this article, we first detail the definition and mathematical properties of \MA gravity, first based on a mathematical intuition obtained by examining the equations (beginning of Section \ref{MAG}) and their relations with Optimal Transport (Section \ref{sec:MAGOT}). Then we re-obtain the same equations from a physical path of reasoning, deriving the laws of motions as the continuum limit of a microscopic model of a large number of particles subject to the principle of large deviations, similar to a least action principle (Section \ref{sec:LDP}). We exploit this physical reasoning to invent a new numerical simulation algorithm (Section \ref{sec:pathbundle}). Finally, we run the same N-body simulation with $512^3$ particles (Section \ref{sect:results}), with Newtonian gravity (using an adpative-mesh algorithm similar to \cite{Couchman1994HydraAA}) and \MAG, both starting from the same $\Lambda$CDM initial conditions, and make some qualitative observations. To our knowledge, this is the first high-resolution simulation of highly non-linear modified gravity, which we believe can be applied to a larger class of modified theories of gravity, such as Gallileon \cite{fairlie-comments}. {\textcolor{black}{For full details of the relation between Monge-Ampère equation and the theories of Galileons, we refer the reader to \cite{albert}.}


\section{Monge-Ampère gravity}
\label{MAG}
In this section, we show how the \MA equation \eqref{eqn:potMA} relates to the usual Poisson equation in Newtonian gravity. We then explain its relation with Optimal Transport. We subsequently use this relation to derive \MAG from a microscopic model and to design a cosmological N-body simulation.
In \MAG \cite{YBGAFA2004,YB_MAG_2011,ambrosio2020mongeampere},
the gravitational force derives from a potential
$\phi$, related to the solution $\Phi$ of a \MA equation:

\begin{equation}
    \det(\D^2\Phi(\br)) = \frac{\rho(\br)}{\rhobar}\,, \\[2mm]
     \label{eqn:potMA}
    \end{equation}
where $\rhobar$ denotes the background density. Next we write
    \begin{equation}
     \Phi(\br) = \frac{\phi(\br)}{4\pi\G\rhobar} + \frac{|\br|^2}{2}\,,     \label{eqn:potMA2}
\end{equation}
and expand the determinant in (\ref{eqn:potMA}) around the identity to obtain the Poisson equation:
\begin{eqnarray}
    1+\trace\left(\D^2\frac{\phi(\br)}{\C}\right) +\mathcal{O}(\phi^2) & = & \frac{\rho(\br)}{\rhobar} \,,
    \label{eqn:MAexpansion}
    \\
 \Delta \phi(\br) & \simeq & 4\pi\G(\rho(\br)-\rhobar)\,.
 \label{eqn:LaplaceMA}
\end{eqnarray}

This corresponds to the usual forms of the \MA \eqref{eqn:MA1} and Poisson equation \eqref{eqn:Poisson1} (using $\Phi$ as defined in eq. \eqref{eqn:potMA} and $\alpha = 1/\rhobar$). In our Poisson equation \eqref{eqn:LaplaceMA}, the average density $\rhobar$ is subtracted out of $\rho(\br)$.
This is a common practice in astrophysics, since a uniform density, $\rhobar$, yields no potential, and is often neglected at the galactic scales whereas taken into account at cosmological scales.

\subsection{Monge-Ampère and optimal transport}
\label{sec:MAGOT}
Optimal Transport concerns the properties of maps $T$ that minimize the integrated weighted quadratic transport cost,
\begin{equation}
\inf\limits_T\left[\int \rho(\br) | \br - T(\br)|^2 d\br \right]   \,,
\label{eqn:TransportCost}
    \end{equation}
\noindent
subject to mass conservation:

 \begin{equation}
    \int_B \rhobar d\bq = \int_{T^{-1}(B)}  \rho(\br)d\br \quad \forall B \,,
    \label{eqn:TransportCnstr1}
\end{equation}
(for all subset $B$, there should be the same quantity of matter in $B$ and in its pre-image through $T$). To analyze the optimization problem above, it is easier to rewrite the constraint in terms of functions :\begin{equation}\int g(\bq)\rhobar d\bq  =
                 \int g(T(\br)) \rho(\br) d\br \quad \forall g.
\label{eqn:TransportCnstr2}
\end{equation}
Intuitively, one can see that the equation above implies \eqref{eqn:TransportCnstr1} if one takes $g(\br)$ = 1 if $\br$ is in $B$ and 0 otherwise. Enforcing a constraint through functions $g(.)$ is a common practice in Finite Elements Analysis (where functions $g(.)$ are referred to as \emph{test functions}).

We shall now follow a derivation similar to the Euler-Arnold formulation of incompressible fluids \citep{AIF_1966__16_1_319_0}, where pressure appears as the Lagrange multiplier associated with incompressibility. We first replace the squared distance in the quantity to be minimized by the dot product:
$$
 \inf_T \left[ \int | \br - T(\br) |^2 d\br \right] \quad \rightarrow \quad
 \sup_T \left[ \int \br \cdot T(\br) d\br \right]
$$
and we inject the constraint as a Lagrange multiplier. Then, our optimization problem \eqref{eqn:TransportCost} subject to mass conservation \eqref{eqn:TransportCnstr2} rewrites as a saddle-point problem:
\begin{equation}
\begin{array}{ll}
          \sup\limits_T \inf\limits_\Psi \scalebox{1.5}{[} & {\cal L}(T,\Psi) =
          \int \rho(\br) T(\br) \cdot \br d \br
          \ +  \\[2mm] & \int \rhobar \Psi(\bq) d\bq\ -
             \int \Psi(T(\br)) \rho(\br) d\br\  \scalebox{1.5}{]}
\end{array}
\end{equation}
where the potential $\Psi$ is the Lagrange multiplier associated with mass conservation \eqref{eqn:TransportCnstr2}.

From the first-order and second order optimality conditions, one deduces that:
\begin{eqnarray}
    \frac{\partial {\cal L}}{\partial T} = 0 & \ \Rightarrow\ &
     \br = \nabla \Psi(T(\br)) \label{eqn:Tpsi} \\[2mm]
    \frac{\partial^2 {\cal L}}{\partial T^2} \ge 0
    & \ \Rightarrow\  & \Psi \mbox{ is a convex function }
\end{eqnarray}

The relation \eqref{eqn:Tpsi} above gives $\br$ as a function of $T(\br)$ (inverse optimal transport map). Since $\Psi$ is convex, the (direct) optimal transport map $T$ is given by the gradient of another convex potential $\Phi$, related with $\Psi$ through its
Legendre-Fenchel transform $\Psi^*$.
\begin{eqnarray}
    T(\br) & = & \nabla \Phi(\br), \quad \mbox{where:} \label{eqn:MongeMA1} \\
    \Phi(\br) & = & \Psi^*(\br) =
    \inf_{\bq} \left[ \bq \cdot \br - \Psi(\bq) \right].
\end{eqnarray}

In thermodynamics, such convex potentials $(\Psi,\Phi)$ are said to be \emph{Legendre dual}. Their duality is a direct consequence of the relation between the Lagrangian and the Hamiltonian formalisms.

Next, $T(\br) = \nabla \Phi$ is inserted into the mass-conservation constraint in Eq. \eqref{eqn:TransportCnstr2}:
\begin{eqnarray}
     \rhobar \int g(\nabla \Phi(\br))|\D^2\Phi(\br)|d\br & = &    \int g(\nabla \Phi(\br)) \rho(\br) d\br \label{eqn:MongeMA2} \\
    \rhobar \det D^2 \Phi & = & \rho(\br).
    \label{eqn:MongeMA3}
\end{eqnarray}

Eq.\eqref{eqn:MongeMA2} is obtained by applying the change-of-variable formula to the LHS of Eq. \eqref{eqn:TransportCnstr2}. This yields the \MA equation \eqref{eqn:MongeMA3} by recalling that Eq. \eqref{eqn:MongeMA2} should be satisfied for all ``test functions $g(.)$. \\

From the above derivation, we deduce the relation between the solution $\Phi$ of the \MA equation \eqref{eqn:potMA}, the optimal transport map $T$ and the gravitational potential $\phi$: 

\begin{itemize}
    \item the optimal transport map is given by $T = \nabla \Phi$;
    \item the gravitational potential $\phi$ is related to $\Phi$ by: $\Phi = \phi/(\C) + |\br|^2/2$,
    because Eq. \eqref{eqn:MongeMA3}
    is identical to the \MA equation \eqref{eqn:potMA}.
\end{itemize}




\subsection{Symmetries of the Monge-Ampère equation}

The Newtonian potential, as the solution of the Poisson equation, is rotation-invariant, that is, if $\Phi$ is a solution of $\Delta \Phi = \rho$, then we have $\Delta (\Phi \circ  R) = \rho \circ R$ for any rotation $R$ (or rigid motion). This explains why a spherical lump of matter (invariant by rotation) remains stable under Newtonian gravity.

The Monge-Ampère equation shows invariance with respect to a wider class of transforms, that preserve volumes (gobally or locally). Intuitively, the Monge-Ampère equation can be thought of as (local) volume conservation for a map $T$ (where the map $T$ is the gradient of a convex potential $\Phi$), hence it is invariant with respect to a volume-conserving transform applied to both sides of the equation. More formally, consider a solution $\Phi$ of the Monge-Ampère equation $|D^2 \Phi| = \rho$, and a transform $\bu: \IR^3 \rightarrow \IR^3$. We shall now see under which condition $\Phi \circ \bu$ is a solution of  $|D^2 [\Phi \circ \bu]| = \rho \circ \bu$. Expanding the l.h.s., one gets:
\begin{eqnarray}
    |D^2[\Phi \circ \bu](\bx)| & = & \left| J\left[ J^t[\Phi \circ \bu](\bx) \right] \right| \nonumber\\
                               & = & \left| J[J\Phi(\bu(\bx)) \times J\bu(\bx) ]^t \right|    \nonumber\\
                               & = & \left| J\left[ J^t(\bu(\bx)) \times J^t\Phi(\bu(\bx)) \right] \right| \nonumber\\
                               & = & \left|\ D^2\bu(\bx) \times J^t\Phi(\bu(\bx)) + \right. \nonumber  \\
                               &   &  \ \       \left.J^t\bu(\bx) \times D^2\Phi(\bu(\bx))\ \ \right|,
                               \label{eqn:composexpansion}
\end{eqnarray}
where $Jf$ denotes the Jacobian matrix of $f$, using the relations $\nabla f = J^t f$ (gradient is a \emph{transposed} Jacobian), $D^2 f = J[J^t f]$, the chain rule and the product rule.

In Equation \ref{eqn:composexpansion}, $D^2\bu$ is a $3\times 3$ tensor with coefficients that are the second order derivatives of $u$. Hence the first term vanishes if $\bu$ is an affine function, in the form $\bu(\bx) = A \bx + \bb$ (or if $u$ has very small second order derivatives), then one gets:
\begin{eqnarray}
    |D^2[\Phi \circ \bu](\bx)| & = & \left| J^t\bu(\bx) \times D^2\Phi(\bu(\bx)) \right| \\
                               & = & \left| J \bu(\bx) \right| \times \left| D^2\Phi(\bu(\bx)) \right|.
\end{eqnarray}
Hence, if $\bu$ is volume-preserving, that is, with unit Jacobian, then the term $|J\bu(\bx)|$ is equal to one, and one gets:
\begin{equation}
    |D^2[\Phi \circ \bu](\bx)| = \left| D^2\Phi(\bu(\bx)) \right|.
\end{equation}

To summarize, if $\Phi$ is a solution of $|D^2\Phi| = \rho$, consider a
function $\bu$ with unit Jacobian and vanishing (or negligible) second order derivatives, then
$| D^2[\Phi \circ \bu]| = \rho \circ \bu$. In other words, the Monge-Ampère equation is invariant
w.r.t. a function that locally behaves like a (volume-preserving) shearing. Clearly, one needs to further analyze the implication for the Monge-Ampère gravitational potential
$\phi$, which will be done in future works. At this point, we have gained the intuition that anisotropic structures (ellipsoids, filaments, \ldots) are more likely to remain stable through Monge-Ampère gravity than with Newtonian gravity. This will be empirically illustrated in the results section.
\begin{figure*}
   \centering
     \includegraphics[width=0.55\columnwidth]{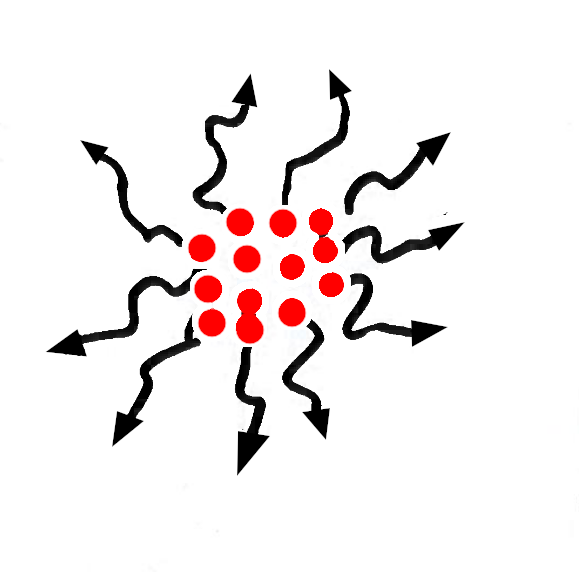}
     \hspace{15mm}
     \includegraphics[width=0.4\columnwidth]{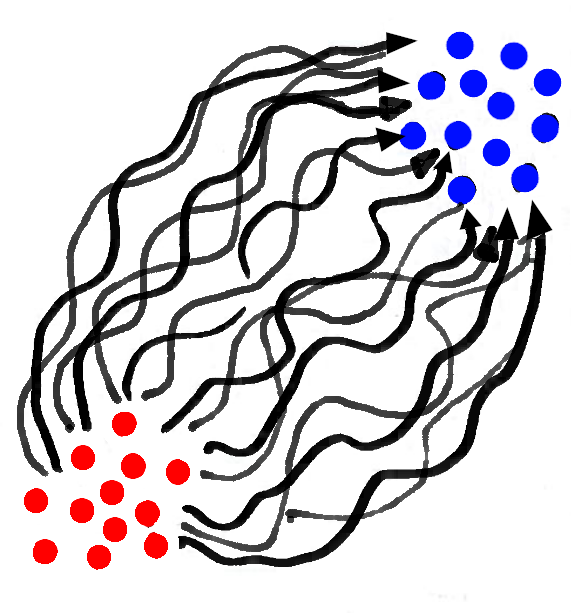}
     \hspace{15mm}
     \includegraphics[width=0.4\columnwidth]{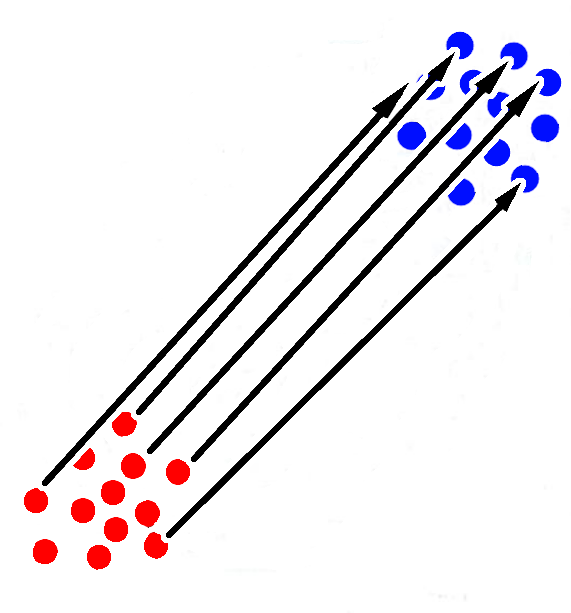}
    \caption{ Left panel: unconditioned motion of $M$ independent Brownian particles; Center panel: motion of independent Brownian particles conditioned by their initial and final positions (in red and blue respectively); Right: conditioned Brownian motion with vanishing noise, all trajectories tend to geodesics.}
    \label{fig:LDP}
\end{figure*}

\subsection{ Monge-Ampère gravity; \\ emergence from the large deviation principle}
\label{sec:LDP}

  The derivation of Monge-Ampère gravity in the beginning of this section is motivated by a mathematical intuition: using the expansion of a determinant around the identity, one can construct an equation similar to Newton-Poisson gravity with additional
  higher-order terms, by replacing the Laplace operator (determinant of the trace of the Hessian) with the Monge-Ampère operator (determinant of the Hessian), as derived in equations (\ref{eqn:potMA}) to (\ref{eqn:LaplaceMA}). \\

  In this section, we detail an alternative derivation, leading to the the very same equation, but this time motivated by a physical principle. We shall use later this alternative physically-motivated path of reasoning to invent an efficient numerical simulation algorithm. \\

  Following \cite{brenier:hal-01137528}, we shall derive \MAG from a microscopic model, that shares some similarities with the relation between optimal transport and Schrödinger bridges \cite{LEONARD20121879}. For the sake of completeness, we mention that there also exists an elegant ``pilot-wave-like derivation \cite{ambrosio2020mongeampere}.

  The idea resembles the least action principle: assume that a motion from a fixed initial condition to a fixed final configuration minimizes (or extremizes) some criterion, deduce the law of motion in-between as a differential relation, then extrapolate it.

So we consider the time evolution of a density field $\rho(\br,t)$, that accounts for a (huge) number of $M$ indistinguishable particles. The only assumption regarding the particles is that they move randomly and do not interact. With this only assumption, everything will spread out and converge to a uniform distribution of particles (see the left panel of Figure \ref{fig:LDP}).

Now we suppose that the density field $\rho(\br,t)$ was observed at time $t=0$ and at a time $t=T$. Then we seek for the ``most probable motion from the initial condition at time $t=0$ (red particles in Figure \ref{fig:LDP}-center panel)
that accounts for the observations at time $t=T$ (blue particles in Figure \ref{fig:LDP}-center panel). As can be seen, this may be thought of as a statistical version of the least action principle, where we seek for the most probable motion (instead of extremizing an action).

The initial condition at time $t=0$ is represented by a permutation $\sigma_0$ of $M$ particles, that accounts for the particle's indistiguishability :
$
\left\{
  X^0_1, \ldots,
  X^0_M \right\} =
\left\{
  \bq_{\sigma_0(1)}, \ldots,
  \bq_{\sigma_0(M)}
\right\}.
$
We suppose that each particle's trajectory is given by $
 \X_i(t) = X^0_i + \epsilon B_i(t)
$ where the $B_i(t)$'s are $M$ independent realizations of a Brownian motion and where $\epsilon$ denotes the amount of noise.

Given a set of points $(Y_i)_{i=1}^M$, 
the probability of the (very unlikely, hence large deviation) event that at time $T$, each position $Y_i$ is occupied by one of the particles $\X_i(t)$ writes:

\begin{equation}
\begin{array}{l}
\Prob\hspace{-2mm}\left( \X_i^\epsilon(T) \underset{\text perm}{\approx} \hspace{-1mm}Y\hspace{-1mm} \right)
\hspace{-1mm}\approx
\frac{(2\pi\epsilon T)^{-\frac{3M}{2}}}{M!}
\hspace{-2mm}\underset{\sigma \in S_M}{\sum}\hspace{-2mm}
\exp\left[
  \frac{- \underset{i}{\sum} | Y_{\sigma(i)} - X_i^0 |^2}
  {2\epsilon T}
\hspace{-1mm}\right]

\end{array}
\label{eqn:LDP}
\end{equation}
where the sum is over all permutations $S_M$ of $[1\ldots M]$ to account for particle's indistinguishability.

Similarly to what happens in path integrals, among the permutations, the one that minimizes
$\underset{i}{\sum} | Y_{\sigma(i)} - X_i^0 |^2$ has a tremendous influence.
As the noise $\epsilon$ tends to zero, the $\log$ of this quantity (a ``smoothed infimum) becomes
the true infimum (Laplace lemma or Boltzmann distribution in statistical physics):

\begin{equation}
\begin{array}{l}
  - \underset{\epsilon \rightarrow 0}{\lim} \, \epsilon \log \Prob
  \left[
  \X_i^\epsilon(T) \underset{\text perm}{\approx} Y
  \right]
  \approx
\underset{\sigma \in S_N}{\inf}
  \left[
     \frac{\underset{i}{\sum}|Y_{\sigma(i)} - X^0_i |^2}{2T}
  \right]
\end{array}
\end{equation}
This corresponds to a discrete version of the optimal transport \eqref{eqn:TransportCost} (the permutation $\sigma$ is the discrete version of the mass-preserving transport $T$).

Moreover, when $\epsilon$ tends to zero, all the trajectories become geodesics (rectilinear uniform), see Figure \ref{fig:LDP}-right:

\begin{equation}
\begin{array}{llcl}
& \X_i(t) & = & X_i^0 + \frac{t}{T} (Y_{(\sigma|Y)(i)}) - X_i^0) \\[2mm]
\mbox{where: } &
(\sigma|Y) & = & \underset{\sigma \in S_N}{\mbox{Arg inf}} \left[
\frac{\sum_i | Y_{\sigma(i)} - X_i^0 |^2}{2T} \right]
\end{array}
\label{eqn:geodesic}
\end{equation}

Then, one can derive a law of motion in the form of an ordinary differential equation by using a simple property: along a uniform rectilinear trajectory (Eq. \ref{eqn:geodesic}), $X^0$ is the closest point -- w.r.t the distance
$\| X - Y \|^2 = \sum_i | X_i - Y_i |^2$ -- to not only $X(T) = (Y_{(\sigma|Y)(i)})_1^N$, but also to any $X(t)$ for
$t$ in $[0,T]$. Hence, one obtains the following relation:

\begin{equation}
  t \frac{d\X_i(t)}{dt} = \X_i(t) - \bq_{(\sigma|\X(t))(i)}
\end{equation}

Let us now see how to derive a force, that is, a second-order equation with respect to time. First, to avoid singularity at time $t=0$, we make a change of time variable $t = \exp(\tau)$:

\begin{equation}
\frac{d \X_i(\tau)}{d \tau} = \X_i(\tau) - \bq_{(\sigma|\X(\tau))(i)}
\label{eqn:tau}
\end{equation}

Now, let us rewrite Eq. \ref{eqn:tau} with the set of all trajectories $\X = (\X_i)_{i=1}^N$. The right hand side can be written as the gradient of a function $\Psi: \IR^{3N} \rightarrow \IR$ of all particle coordinates:
\begin{equation}
\begin{array}{llcl}
& \frac{d \X_i(\tau)}{d \tau} & = & - \frac{\partial \PPhi}{\partial \bx_i}(\X_1(\tau), \ldots, \X_N(\tau)) \\[2mm]
\mbox{where} &
\PPhi(\bx_1, .. \bx_N) & = & \underset{\sigma \in S_N}{\inf} \left[
\frac{\sum_i | \bx_i - \bq_{\sigma(i)} |^2}{2}
\right]
\end{array}
\label{eqn:speed}
\end{equation}

To derive an equation that resembles Newton's law, let us now compute the second order derivative (acceleration). Equation \ref{eqn:speed} enjoys a very particular property. Its right hand side remains the same when differentiating with respect to time:

\begin{equation}
\begin{array}{lcl}
   \frac{d^2 \X(\tau)}{d\tau^2}
        & = & - \D^2\left(\PPhi(\X(\tau))\right) \frac{d\X(\tau)}{d\tau} \\[2mm]
        & = & \D^2\left(\PPhi(\X(\tau)\right) \nabla \PPhi(\X(\tau)) \\[2mm]
        & = & \nabla \left( \frac{\| \nabla \PPhi(\X(\tau)) \|^2}{2}\right)\\[2mm]
        & = & -\nabla \PPhi(\X(\tau))
\end{array}
\label{eqn:accel}
\end{equation}

Finally, from the definition of the potential $\PPhi$ (Eq. \ref{eqn:speed}), it is easy to check that the acceleration $d^2\X_i(\tau)/d\tau^2$ of one of the particles is given by:
\begin{equation}
    \begin{array}{lcl}
    \frac{d^2\X_i(\tau)}{d\tau^2} & = &
    -\frac{\partial \PPhi}{\partial \X_i} = \X_i - \bq_{(\sigma | \X(\tau))(i)} \\[2mm]
    & = & \X(\tau) - \bq_{(\sigma | \X(\tau))(i)} \\[2mm]
                                & = & \X(\tau) - \nabla \Phi(\X(t)) \\[2mm]
                                & = & -\nabla \phi(\X(\tau)).
    \end{array}
\end{equation}
In the last two lines, we consider the continuum limit, where the number $M$ of particles tends to infinity, and write the equations for the potentials $\phi$ and $\Phi$,  where $\Phi$ is the solution of the Monge-Ampère equation, and $\phi$ is determined by $\Phi(\bx) = \bx^2/2-\phi(\bx)$. \\

To summarize the derivations above:
\begin{enumerate}
\item From the principle of large deviations conditioned by both the initial condition at $\tau=\tau_1$ and the observation at time $\tau=\tau_2$, one deduces that between $\tau=\tau_1$ and $\tau=\tau_2$, each particle is subject to a force.
\item This force is equivalent to a ``repulsion by a ``virtual particle, that corresponds to the particle's image through optimal transport towards the initial condition. One may be surprised that it is a ``repulsive force, and that only a single particle is involved, but we recall that this ``virtual particle accounts for the influence of all real particles through optimal transport.
\item This force derives from a potential, that is obtained as the solution of the Monge-Ampère equation
$\det(D^2(\Phi)) = 4\pi\G\rho$, similar to the usual Poisson equation, but with additional non-linear terms (the off-diagonal coefficients of the Hessian).
\item The \MA gravitational force is given by:
\begin{eqnarray}
     \frac{d^2\X_i(\tau)}{d\tau^2}      & = & \X(\tau) - \bq_{(\sigma | \X(\tau))(i)}
     \label{eqn:MA_force_discrete}
     \\
                                & = & \X(\tau) - \nabla \Phi(\X(t))
                                \ =\ -\nabla \phi(\X(\tau))
     \label{eqn:MA_force_continuous}
\end{eqnarray}
where the continuous version \eqref{eqn:MA_force_continuous} uses the relation $\bq = \nabla \Phi$, with $\Phi$ the solution of the MA equation (see Equations \ref{eqn:MongeMA1} to \ref{eqn:MongeMA3}).
\end{enumerate}

One can gain more intuition about the repulsion from the ``virtual particle that corresponds to the optimal-transport image.
Starting from the initial condition, and with particles that move according to $M$ independent Brownian motions, the principle of large deviation considers the probability of a very unlikely event: exactly one particle is observed at each position $(Y_i)_{i=1}^M$. It is known that moving along the direction of optimal transport tends to regularize the distribution of the particles \cite{DBLP:journals/tog/XinLCCYTW16} and to maximize entropy. In our case, moving along the \emph{opposite} direction favors the apparition of structures, by directly leading to the (very unlikely) event.




\section{Cosmological simulation
        of Monge-Ampère gravity
     }
     \label{MAGsim}

\begin{figure}
    \centering
    \includegraphics[width=0.8\columnwidth]{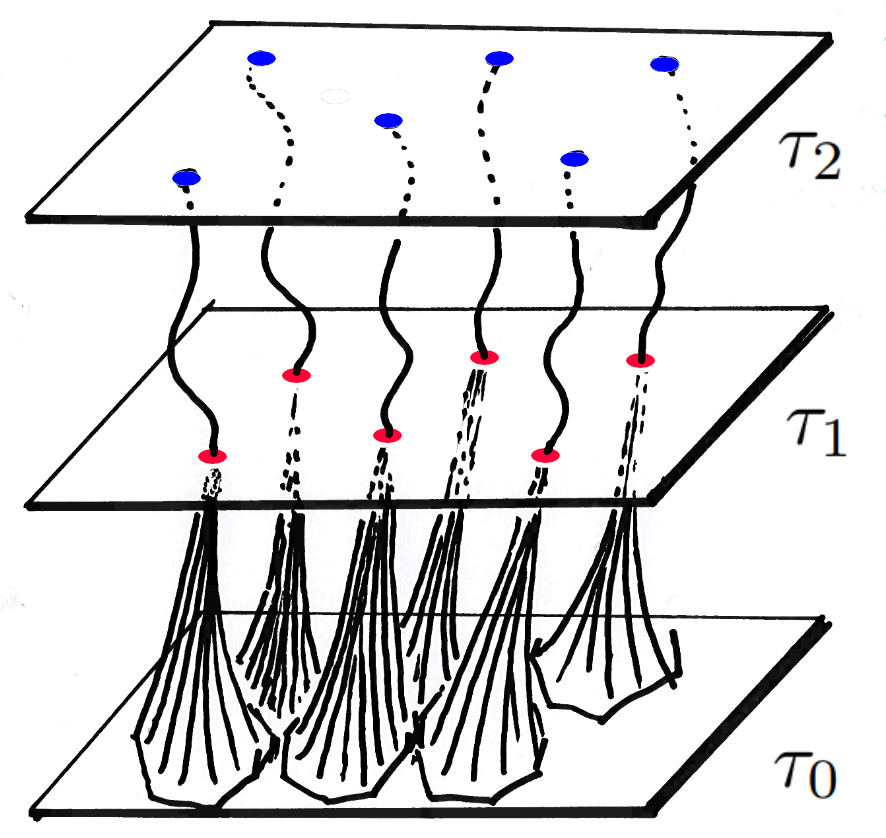}
    \caption{Semi-discrete simulation of Monge-Ampère gravity with the path bundle method. To derive the force, we suppose that at a time $\tau_0$ matter takes the form of a uniform density field. Between times $\tau_0$ and $\tau_1$, matter clusters into $N$ structures, symbolized as red points. In addition, we suppose that the structures are observed at certain locations at a later time $\tau_2$. From the so-conditioned large deviation principle, one deduces the law of motion of the $N$ structures}
    \label{fig:pathbundle}
\end{figure}

In this section, we detail our numerical simulation algorithm for Monge-Ampère gravity. Our goal is now to simulate a set of $N$ objects (galaxies), represented by points, or ``macro-particles $(\br_i)$ with masses $(m_i)$, with initial positions $\bq_i = \br_i(\tau_1)$ and velocities $\bv_i(\tau_1)$. \\

\begin{figure*}
    \includegraphics[width=\textwidth]{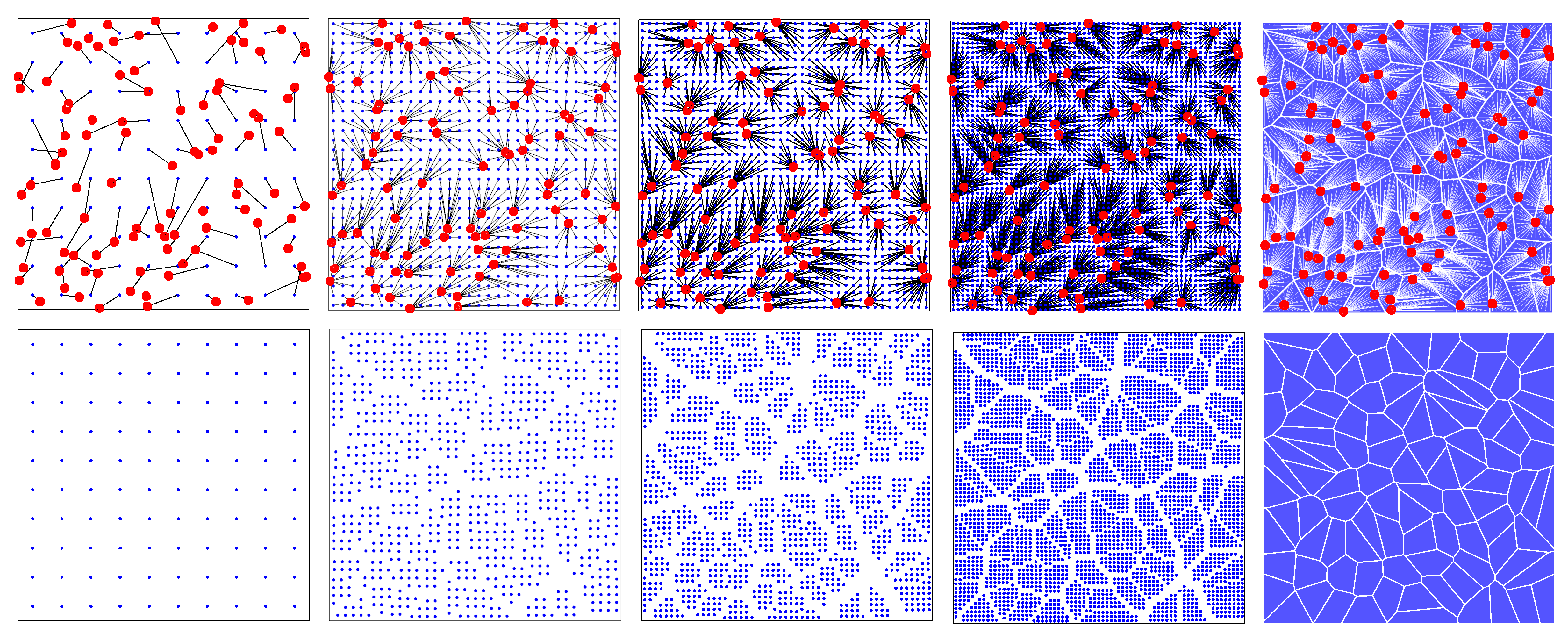}
    \caption{Semi-discrete optimal transport as the continuum limit of discrete optimal transport. First four columns: each macro-particle is associated with 1,9,16,25 microparticles respectively. Rightmost column: when the number of microparticles tends to infinity, each macroparticle is associated with a polygonal region, called a Laguerre cell, computed explicitly by our algorithm.}
    \label{fig:from_d_to_sd}
\end{figure*}

Clearly one could use the discrete \MA gravitational force \eqref{eqn:MA_force_discrete} to derive a N-body simulation, with $F_i = \br_i - \bq_{(\sigma|\X)(i)}$, where $(\sigma|\X)(i)$ is the optimal permutation
of the $N$ points that minimizes the (weighted) transport cost
$C(\sigma) = \sum_i m_i | \br_i - \bq_{\sigma(i)} |^2$.
However, finding $(\sigma|\X)(i)$ at each time-step is a very costly operation, typically in $O(N^2\log(N))$ \cite{Bertsekas1989}. This algorithmic complexity prevents us from using discrete optimal transport in cosmological simulations, for which we need at least $512^3$ particles to have sufficient resolution to see the evolution of structures. For this reason, as explained in the next subsection, we use instead the \emph{continuous} \MA force \eqref{eqn:MA_force_continuous}, leading to a much more efficient numerical solution mechanism that exploits the structure of the problem (convexity and smoothness). In practice, complexity becomes $O(N\log(N)$ \cite{levy_mnras_2021} (to be compared with $O(N^2\log(N)$).

\subsection{The path bundle method}
\label{sec:pathbundle}
We shall now derive the laws of motion of the $N$ macroscopic particles based on the
continuous \MA force. To do so, we consider that each of the $N$ macro-particles
is composed of a huge number of micro-particles, with a total number $M >> N$ of micro-particles, then we will consider the continuum limit. For now, in the discrete setting, the mass $m_i$ of a macro-particle $i$ may be thought of as the number of micro-particles in $i$.

The $M$ indistinguishable micro-particles ($M >> N$) move according to $M$ independent Brownian motions conditioned as shown in Figure \ref{fig:pathbundle}: we suppose that the matter that corresponds to each particle $(m_i,\br_i)$ at the initial condition $\tau_1$ has lumped from a uniform distribution of matter, at a ``pre-initial condition $\tau = \tau_0$. This uniform pre-initial condition may be thought of as a regular grid of micro-particles.

In addition, we suppose that at time
$\tau_1$, all the micro-particles that clustered into a macro-particle have the same velocity. Hence, from time $\tau_1$ and later, the $N$ structures remain intact and can be modeled as $N$ macro-particles. In other words, the trajectories of the microscopic particles that clustered into a structure form a ``bundle, that merges at time $\tau_1$. Since microscopic particle's velocities match at time $\tau_1$, between times $\tau_1$ and $\tau_2$, each ``bundle is tight, and appears as a unique trajectory of a macroscopic particle.

From the large deviation principle detailed in the previous section,
one deduces that each macroscopic particle is subject to a force that results from the combined influence of all the trajectories of the micro-particles.

Now remember the intuition that a particle is ``pushed away by its optimal-transport image (equation \ref{eqn:MA_force_discrete}, end of previous Section). In the present context, at a
time $\tau$ between $\tau_1$ and $\tau_2$, we consider that a macro-particle is
composed of a huge number of micro-particles. Each of these micro-particle is ``pushed away by its optimal-transport image. The force that the macro-particle is subject to corresponds to the sum of each of its corresponding micro-particle's contribution.

As shown in Figure \ref{fig:from_d_to_sd}, taking the continuum limit, the regular grid of micro-particle at the pre-initial condition at $\tau = \tau_0$ converges to the uniform density $\rho(\bq,\tau_0) = \bar{\rho}$, and the force influencing each macro particle
can be derived from the optimal transport between the uniform density and the set of (weighted) particles $(m_i,\br_i)$. Thus, the gravitational force $F_i$ for a given object $\br_i(\tau)$ writes:
\begin{equation}
    F_i(\tau) = -\nabla \phi(\tau) = \br_i - \nabla \Phi(\br_i,\tau) \quad \forall \tau \ge \tau_1
\end{equation}
where $\Phi(.,\tau)$ is the solution of the \MA equation $\det(D^2\Phi(.)) = \rho(.,\tau)$. \\

Note that the density field $\rho(.,\tau)$ as well as the
potentials $\phi$ and $\Phi$ are concentrated on the $N$ punctual masses $\br_i(\tau)$, hence one needs to define the differential operators $\nabla$ and $\det(D^2(.))$ in a generalized sense.
In the so-called \emph{semi-discrete} setting, the gradient $\nabla \phi$ is replaced with the subdifferential, that associates a region of space $V_i(\phi)$ to each point $\br_i$ \cite{BrenierPFMR91,DBLP:journals/cgf/Merigot11,journals/M2AN/LevyNAL15,levy_mnras_2021,vhauss_prl_2022,nikak_prl_2022},

\begin{equation}
  \begin{array}{l}
    \nabla \phi = V_i(\phi) = \\[2mm]
    \{ \bq \ ; \ | \bq - \br_i |^2 - \phi_i
      \le | \bq - \br_j |^2 - \phi_j \quad \forall 1 \le j \le N \}\,
   \end{array}
   \label{eqn:Laguerre}
\end{equation}
\noindent
where the so-defined $V_i$ regions (Laguerre cells) form a partition of $V$ (Laguerre diagram), parameterized by the $N$ positions $\br_i(\tau)$ of the points and by the potentials $\phi_i(\tau) = \phi(\br_i(\tau))$. Put differently, each Laguerre cell $V_i$ corresponds to the continuum of optimal-transport images that ``pushes the particle $i$ (rightmost column of Figure \ref{fig:from_d_to_sd}).

Integrating the influences of all the points $\bq$ in $V_i$, one obtains the expression of the gravitational force:
\begin{equation}
   \frac{d^2 \br_i(\tau)}{d \tau^2} = F_i(\tau) = \br_i(\tau) - \bg_i(\tau)
   \label{eqn:MAForce}
\end{equation}
where $\bg_i$ denotes the barycenter of the Laguerre cell $V_i$. The Laguerre cells are determined by the points $(\br_i)$ and the potentials $(\phi_i)$ that realize the optimal transport of the $(\br_i)$'s to the uniform density. How to compute the potentials
$(\phi_i)$ is explained later, in Section \ref{sec:solve}.


As compared to the fully discrete setting considered at the beginning of this section, each point $\br_i$ is repelled by the barycenter of the region of space that corresponds to its image through optimal transport (whereas in the discrete setting, it was repelled by a single point $\bq_{\sigma(i)}$). Not only this gives more spatial accuracy for the force, but also it makes it possible to replace the combinatorial search for $(\sigma|\X)$ with a much more efficient smooth and convex optimization.

\begin{figure*}
    \includegraphics[width=\textwidth]{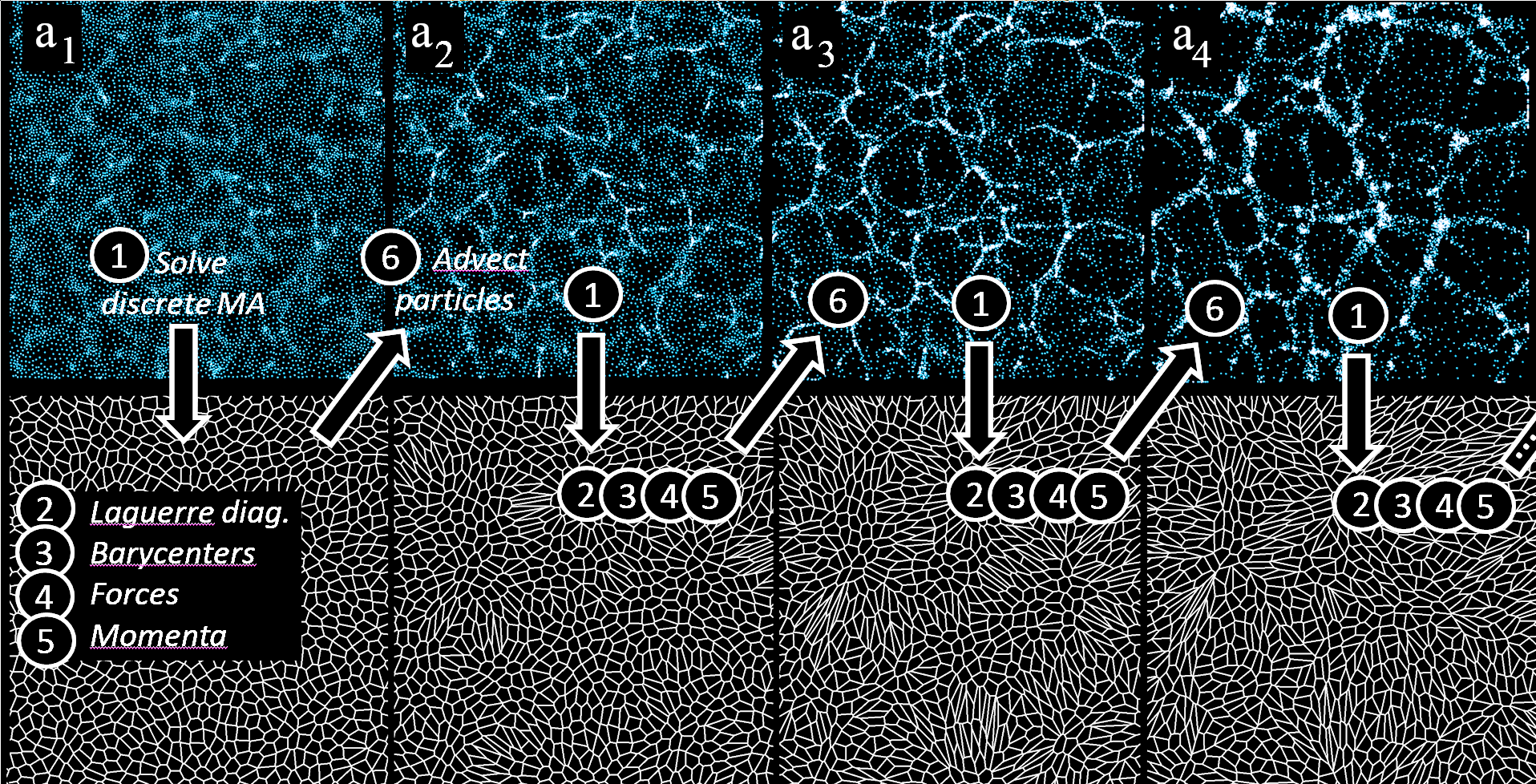}
    \caption{Four timesteps $a_1$, $a_2$, $a_3$, $a_4$ of a 2D cosmological Monge-Ampère gravity simulation, using the expansion scale factor $a$ as the time variable (3D examples shown P. \pageref{fig:hydra16M}). Each timesteps consists of the following substeps: (1) solve the discrete Monge-Ampère equation to compute the potentials $\phi_i$; (2) construct the Laguerre diagram; (3) compute the cell barycenters; (4) compute the forces (5) update the momenta; (6) advect the particles. The bottom row shows a zoom on the Laguerre cells of the central zone.
    }
    \label{fig:BMAG_2d}
\end{figure*}

\subsection{Cosmological simulation}
\label{sec:cosmo}
Based on the expression of the Monge-Ampère gravitational force above, we shall now detail our cosmological simulation algorithm. First, we recast the equations of motion in cosmological time and space coordinates, using the expansion scale factor $a$ as the time variable, and co-moving coordinates $\bx$ defined by $\br=a \bx$:
\begin{equation}
    \frac{d{\bf x}}{da}=\frac{{\bf p}}{m a^2 S(a)} \quad ; \quad \frac{d{\bf p}}{da}=\frac{1}{a S(a)}\left[a^2 {\bf F_{\rm g}} + \frac{m H_0^2\Omega_{\rm m}}{2}{\bf x}\right]
    \label{eqn:comoving}
\end{equation}
where $\bF_g$ denotes the gravitational force (Newtonian or Monge-Ampère) and
$S(a)= a H_0 (\Omega_{\rm m} a^{-3} + (1 - \Omega_{\rm m} - \Omega_{\rm \Lambda} )a^{-2}+ \Omega_{\rm \Lambda})^{1/2},
$
where $\bv = a\bu$ denotes the peculiar velocities, ${\bf p}= m a {\bf v}$ momenta, $H_0$ the Hubble constant, $\Omega_{\rm m}$ the matter density and $\Omega_{\rm \Lambda}$ the cosmological constant. As we assume periodic boundaries for our cosmological boxes, the second term in the square bracket vanishes. For this work, our simulations assume a FLRW background.\\

Our simulation algorithm is very similar to a classical N-body simulation with semi-implicit Euler time integration, the only difference is the definition of the gravitational force ${\bf F_{\rm g}}$. \\

Hence, one timestep of our simulation consists of the following operations, summarized in Figure \ref{fig:BMAG_2d}:
$$
 \begin{array}{ll}
 \ (1): & \mbox{ Solve for the potentials } (\phi_i) \mbox{ using the points } (\bx_i) \\ \ (2): & \mbox{ Compute the Laguerre cells } \{V_i(\bx_i,\phi)\} \\
 \ (3): & \mbox{ Compute the barycenters } (\bg_i) \mbox{ of the } \{V_i\} \mbox{ cells } \\
 \ (4): & \mbox{ Compute the forces } \bF_{MA} \mbox{ using Eq. \ref{eqn:MAForce} } \\
 \ (5): & \mbox{ Update momenta: } \bp \leftarrow \bp + \frac{\delta a}{S(a)} a {\bf F}_{\rm MA} \\
 \ (6): & \mbox{ Update positions: }
 \bx \leftarrow \bx  + \frac{\delta a}{m a^2 S(a)}\bp \\
 \ (7): & \mbox{ Update time: } a \leftarrow a + \delta a
 \end{array}
$$

\begin{itemize}
\item Step (1) solves for the potential that corresponds to the optimal transport between the set of points $(m_i,\bx_i)$ and the uniform density $\rhobar$, more details are given below.

\item Steps (2),(3),(4) direcly use the Laguerre diagrams computed by the KMT-Newton algorithm. Our algorithm shares some similarities with the moving-mesh algorithm in AREPO  \cite{10.1111/j.1365-2966.2009.15715.x}, in the sense that it also uses a dynamic unstructured polyhedral mesh. Besides the difference that here we are simulating a modified gravity, the main differences are that here we use the Laguerre cells instead of Voronoi cells in AREPO and in addition our scheme is purely Lagrangian with exact volume conservation whereas AREPO is based on an Arbitrary Lagrangian Eulerian scheme.

\item Steps (5) and (6) correspond to a semi-implicit Euler scheme with the scale factor $a$ used as a time variable, where $\delta a$ denotes the time-step.

\item Step (7) updates the time parameter $a$. In addition, at this step, we check the Courant-Friedrich-Lewy (CFL) condition \citep{CFL}. This condition, required to ensure the stability of the numerical simulation, imposes an upper bound for the timestep $\delta a$. In our case, the CFL condition imposes that during a single timestep, a particle should not travel further away than the size of its Laguerre cell (see
\cite{doi:10.1137/16M106666X} for the derivation of a CFL condition in a similar context). If the CFL condition is not satisfied by the timestep $\delta a$, a smaller timestep $\delta a'$ is computed in such a way that the fastest particle meets this condition, and several substeps $a \leftarrow a + \delta a'$ are computed until $a$ reaches $a + \delta a$.
\end{itemize}

\begin{figure*}
    \centering
    \includegraphics[width=\textwidth]{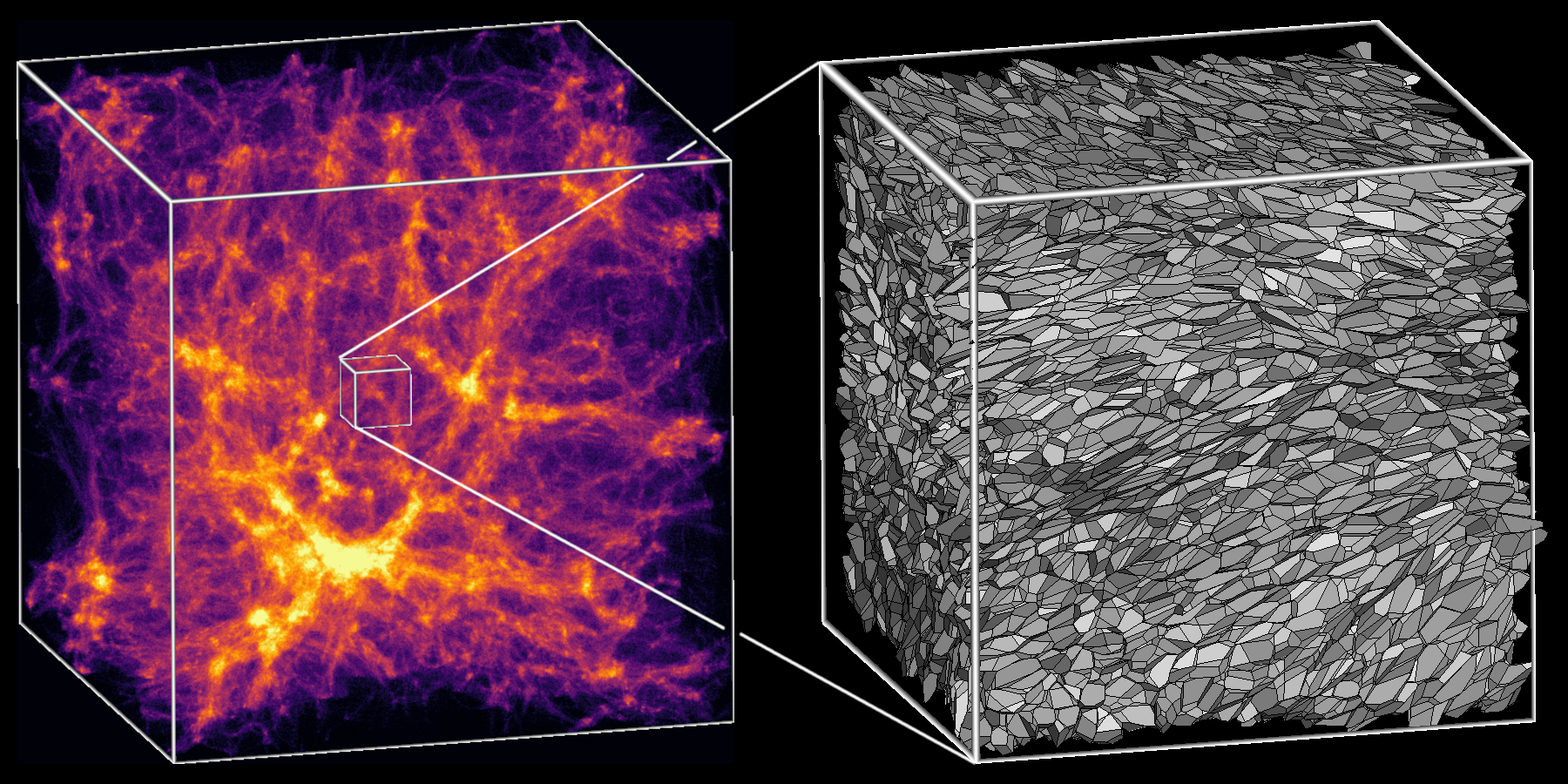}
    \caption{Simulation of Monge-Ampère gravity (60 Mpc/h, $256^3$ particles), and zoom on the Laguerre cells of the central region.}
    \label{fig:laguerre_mesh}
\end{figure*}

\subsection{Solving the semi-discrete Monge-Ampère equation}

\label{sec:solve}

In the N-body simulation algorithm above,
step (1) requires to solve for the gravitational potential $\phi$ in
the Monge-Ampère equation (\ref{eqn:MongeMA3}) recalled below:

$$
  \det(\Phi) = \det D^2\left( \frac{1}{4\pi\G\bar{\rho}} \phi + \frac{\bx^2}{2} \right) = \frac{\rho}{\bar{\rho}},
$$
which is a non-linear equation. To solve it numerically, we use a Newton method, that solves a series of linear equations that converges to the solution.

We first give an intuition of the method using a continuous-setting derivation (with a continuous density field $\rho$), then we shall see how to adapt it to our semi-discrete setting (with individual point masses $\bx_i$,$m_i$).

In the continuous setting, we use the following order-1 expansion of the
Monge-Ampère operator $\det D^2(.)$ in function of a small update $\delta \phi$ of the potential:
\begin{equation}
\det D^2\left(\Phi + \frac{\delta \phi}{4 \pi \G \bar{\rho}}\right) \simeq \det D^2(\Phi) + \frac{1}{4\pi\G\bar{\rho}} \Delta (\delta \phi).
\label{eqn:deltaPhiExpansion}
\end{equation}

Plugging (\ref{eqn:deltaPhiExpansion}) into the Monge-Ampère equation, one gets the equations that
characterizes the update $\delta \phi$ to be applied at each Newton step:
$$
\det D^2 \Phi + \frac{1}{4\pi\G\bar{\rho}} \Delta \delta \phi \simeq \frac{\rho}{\bar{\rho}}$$
or:
\begin{equation}
\Delta \delta \phi \simeq 4 \pi \G \left( \rho - \bar{\rho} \det D^2 \Phi \right).
\label{eqn:discreteMA}
\end{equation}

Putting everything together, one obtains the following Newton algorithm:

\begin{equation}
\begin{array}{ll}
\ (0): \phi \leftarrow 0 \\
\ (1): \mbox{while } \| \rho - \bar{\rho} \det(D^2 \Phi) \|^* < \epsilon \\
\ (2): \quad \mbox{solve for } \delta \phi \mbox{ in }
    \Delta(\delta \phi) = \rho - \bar{\rho} \det(D^2 \Phi) \\
\ (3): \quad \mbox{find descent parameter } \alpha \\
\ (4): \quad \phi \leftarrow \phi + \alpha \delta \phi
\end{array}
\label{eqn:MAalgo}
\end{equation}

We give now more details about the algorithm above:
\begin{itemize}
\item in step (0), the potential $\phi$ is set to 0, which means $\Phi(\br) = \br^2/2$, and the transport $T = \nabla \Phi$ is initialized with the identity. The next steps will update $\phi$, which will ``morph $T$ from the identity to the optimal transport;
\item in step (1), $\|.\|^*$ denotes some function norm and $\epsilon$ a convergence threshold (more on this later);

\item in step (2), one solves a Poisson equation to find $\delta \phi$
(one can drop the $4\pi\G$ factor since $\delta \phi$ is used as a \emph{direction} of descent). For a smooth density $\rho(.)$, in periodic space, one could use the Fourier transform to solve the Poisson equation, as suggested in \cite{ZHELIGOVSKY20105043}. However,
as mentioned in the previous section, in our case, the density $\rho$ is singular, and matter is concentrated on a discrete set of points $(\bx_i)_{i=1}^N$ with masses $(m_i)_{i=1}^N$. The potential $\phi$ becomes a vector $(\phi_i)_{i=1}^N$ of values attached to the individual points $\bx_i$. As in the previous section, we need to consider the differential operators acting on $\phi$ in a general sense (subdifferentials), hence  Equation (\ref{eqn:discreteMA}) becomes:
\begin{equation}
   \hat{\Delta} \phi = m_i - \bar{\rho} |V_i(\phi) |
   \label{eqn:DiscretePoisson}
\end{equation}
where $|V_i(\phi)|$ denotes the volume of the Laguerre cell associated with point $\bx_i$ given the
vector of potentials $(\phi_i)$, as defined in Equation \ref{eqn:Laguerre}. Figure \ref{fig:laguerre_mesh} shows what these Laguerre cells look like in a cosmological simulation. In Equation \ref{eqn:DiscretePoisson}, the $N \times N$ matrix $\hat{\Delta}$ denotes the $\mathbb{P}_1$ Laplacian, that is, the classical
Finite Element discretization of the Laplacian projected onto a basis of piecewise linear functions. The  coefficients $c_{ij}$ of the discrete Laplacian are given by:

\begin{equation}
\begin{array}{lcl}
 c_{ij} & = & \bar{\rho} \frac{1}{2} \frac{1}{\| \bx_i - \bx_j \|} \left| V_i(\phi) \cap V_j(\phi) \right| \quad \forall i \neq j \\[5mm]
 c_{ii} & = & -\sum_{j \neq i} c_{ij},
\end{array}
\end{equation}
where $|V_i(\phi) \cap V_j(\phi)|$ is the area of the common face shared by the Voronoi cells
$V_i(\phi)$ and $V_j(\phi)$ (or $\emptyset$ is there is not such a face). \\

More formally, as detailed in \cite{journals/M2AN/LevyNAL15,DBLP:journals/corr/KitagawaMT16,DBLP:journals/cg/LevyS18}, one can obtain the same equation by maximizing the Kantorovich dual:
$$
 \quad \quad K(\phi) = \rhobar \sum_i \int_{V_i(\phi)} \left( | \bq - \bx_i |^2 - \phi_i \right) d\bq + \sum_i m_i \phi_i,
$$ then the right hand side of the discrete Poisson equation (\ref{eqn:DiscretePoisson}) is obtained from the negated gradient of $K(.)$,
and the discrete Laplacian $\hat{\Delta}$ is obtained from the Hessian matrix
of $K(.)$. Interestingly, this equation also gives a physical meaning to the stopping criterion for step (1): one can see that the gradient of $K(.)$ -- or the right hand side of
the discrete Poisson equation -- correspond to the mass errors (desired mass $m_i$ minus
transported mass $\bar{\rho} | V_i(\phi) |$). We chose to stop Newton iterations as soon as the
error of the worst particle is smaller than 1\% of the particle's mass. In other words,
we use the $L_\infty$ norm for $\|.\|^*$ in step (1).

\item in step(3), the descent parameter $\alpha$ is determined. We use the KMT criterion \cite{DBLP:journals/corr/KitagawaMT16}, that provably generates a sequence of iterates that converges to the solution of the semi-discrete Monge-Ampère equation: the descent parameter $\alpha$ is decreased until the volume of the smallest Laguerre cell is larger than a certain threshold, $\frac{1}{2}\left( \inf_i |V_i(0)|, \inf_i(m_i) \right)$, where $V_i(0)$ corresponds to the Laguerre cell associated with point $\bx_i$ with zero potential (in other words, $\bx_i$'s Voronoi cell).

\end{itemize}

The 3D version of the algorithm above is fully detailed in \cite{journals/M2AN/LevyNAL15}, and used in the context of early Universe reconstruction \cite{levy_mnras_2021,vhauss_prl_2022,nikak_prl_2022,Farnik2024arXiv}.
In early Universe reconstruction, a single instance of the Monge-Ampère equation needs to be solved, but in our case, we need to solve a Monge-Ampère equation \emph{at each timestep}, and solving this highly non-linear equation is much more costly than solving a Poisson equation involved in classical N-body simulations:
in algorithm (\ref{eqn:MAalgo}), one can see that \emph{a series of Poisson equations} are solved at each timestep.
However, for Monge-Ampère, the CFL condition is favorable, and authorizes much larger timesteps than with classical N-body simulations (typically 20 times larger). Even with this observation, running times remain much longer than a classical N-body simulation, because each Monge-Ampère timestep takes around 40 times longer than a classical Poisson solve or FMM time (in other words, we gain a 10x factor and loose a 40x factor, the net factor is around 4x slower). \\

\begin{figure}
    \centering
    \includegraphics[width=\columnwidth]{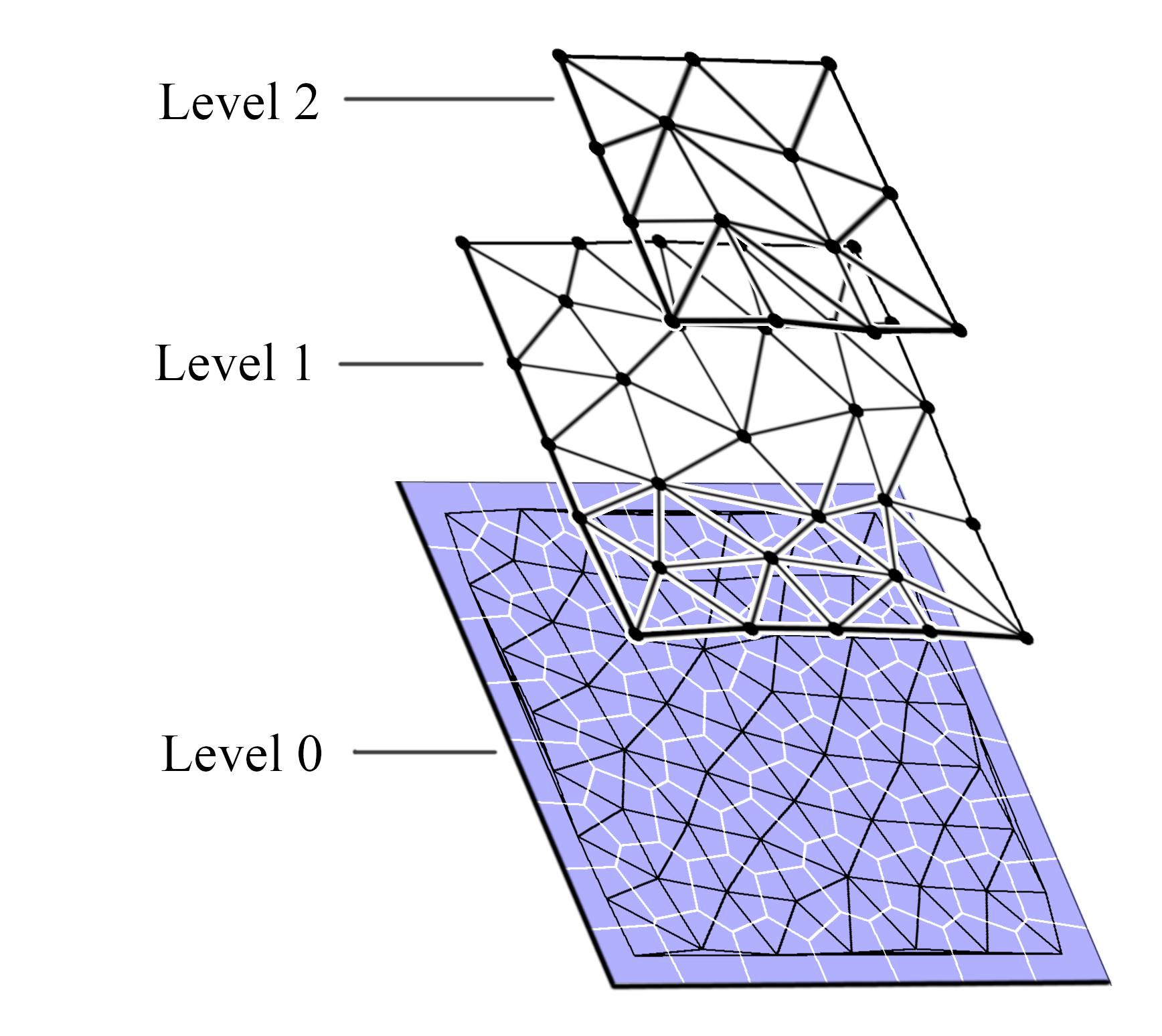}
    \caption{The discrete Laplacian matrix seen as a hierarchy of graphs: at level 0, Laguerre cells are shown as white lines, and the dual graph as thick black lines. To solve the discrete Poisson equations efficiently, the algebraic multigrid method creates a hierarchy of coarser versions of the dual graph (levels 1, 2 \ldots) by lumping values from finer level's nodes.}
    \label{fig:primal_dual}
\end{figure}

To improve computation time, we take a closer look at the discrete Poisson equation (\ref{eqn:DiscretePoisson}) and revisit the way to solve it: in our previous works \cite{levy_mnras_2021,vhauss_prl_2022,nikak_prl_2022}, we used a Jacobi-preconditioned conjugate gradient implemented on the GPU \cite{DBLP:journals/paapp/BuatoisCL09}. If our density fields were discretized on regular grids, it is well known
that our Poisson equations could be solved in linear time (see \cite{ipol.2018.228} and references herein), using a multigrid method. Seen in frequency space, such methods may be thought of as solving for each frequency of the solution on a grid with a resolution well adapted to the frequency. However, we  cannot use such a multigrid method because our discretization is supported by the cells of the Laguerre diagram, that is, an \emph{irregular} mesh (white cells in Figure \ref{fig:primal_dual}). However, there exists a class of multigrid methods, called \emph{algebraic multigrid methods} (AMG), that solely manipulate the matrix of the system, and that create a hierarchy of operators by analyzing the sparsity pattern of the matrix, seen as a graph: the discrete Laplacian matrix $\hat{\Delta}$ has a non-zero coefficient $c_{ij}$ each time the Laguerre cells of $\bx_i$ and $\bx_j$ touch along a common face. This defines a graph, shown in black in Figure \ref{fig:primal_dual}. The algebraic multigrid method creates a hierarchy of matrices, by merging some vertices of this graph, and lumping the associated coefficients accordingly. This hierarchy of matrices play the same role as the staggered grids in multigrid methods, and treat each frequency with a well-adapted resolution. We used the implementation in \cite{Demidov2019}, that uses this strategy to create a preconditioner, and that is especially well adapted to our context, with a large-scale Poisson system supported by an irregular grid. The overall acceleration factor as compared to our previous implementation is between 1.5x and 2x faster.

\begin{figure*}
    \centering
    \includegraphics[width=\textwidth]{MA_3D_z1_300mpch.png}
    \caption{\textcolor{black}{Simulation of Monge-Ampère gravity in a cube of 300 Mpc/h edge length with $512^3$ particles (100 millions). This image: timestep at z=1, 3d field and 2D cross-section. Other timesteps and comparisons with $\Lambda$CDM are shown in Figure \ref{fig:300Mpch}.}}
    \label{fig:MA_3D}
\end{figure*}

\begin{figure*}
    \colorbox{black}{
    \begin{minipage}{0.97\textwidth}
    \centerline{
      \includegraphics[width=0.42\textwidth]{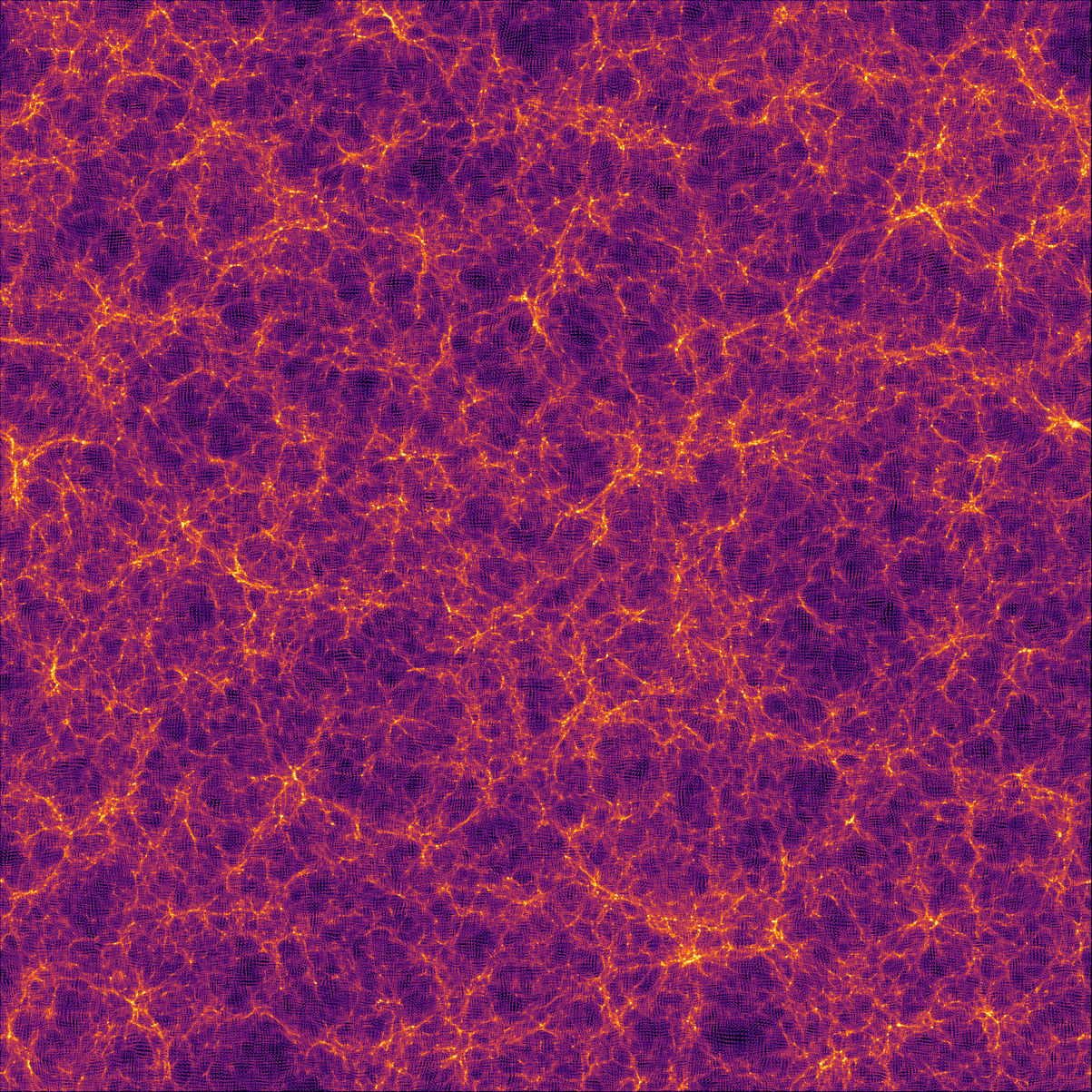}
      \hspace{10mm}
      \includegraphics[width=0.42\textwidth]{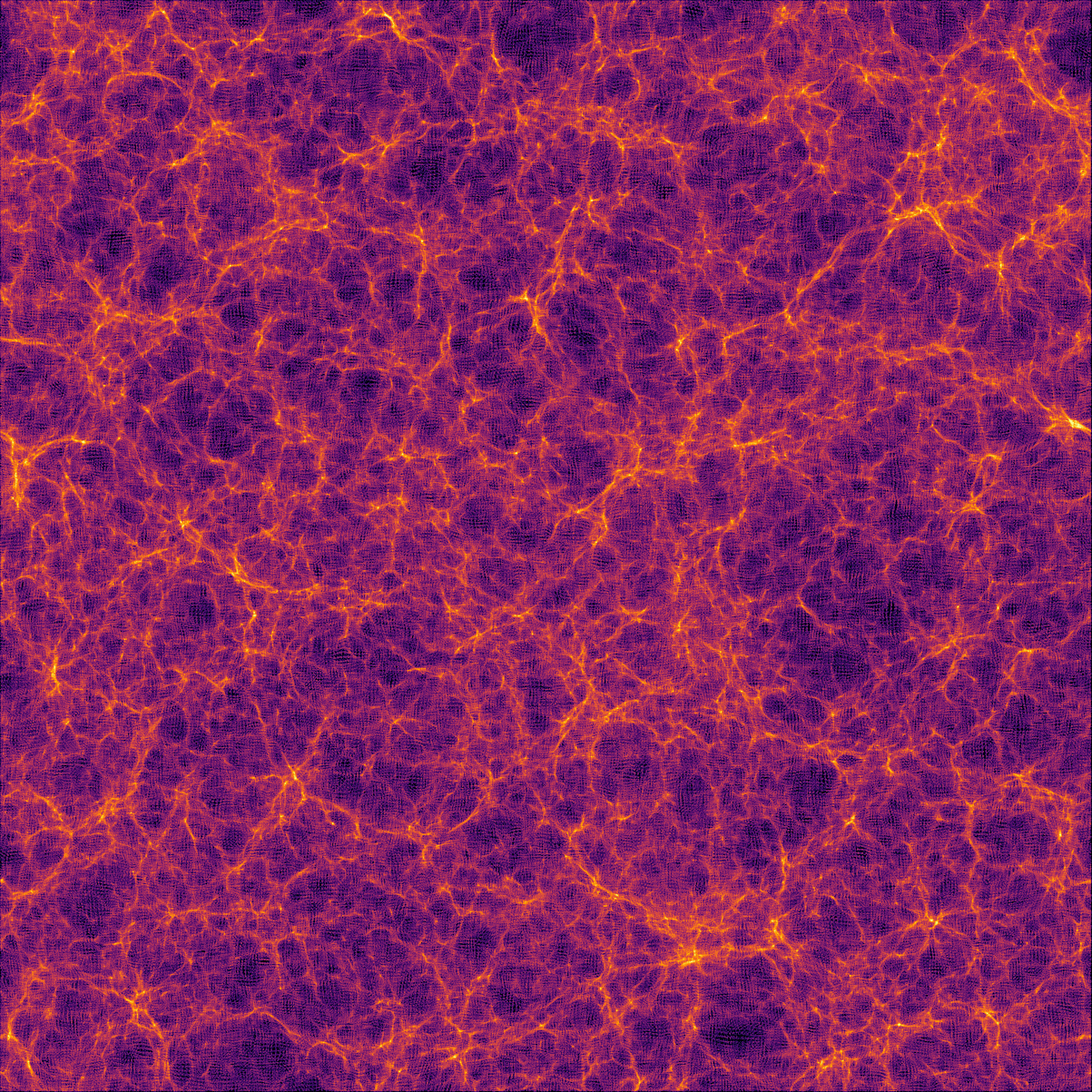}
    }
    \centerline{\color{white} $\Lambda$CDM, $z=5$ \hspace{55mm}Monge-Ampère, $z=5$}
    \centerline{
      \includegraphics[width=0.42\textwidth]{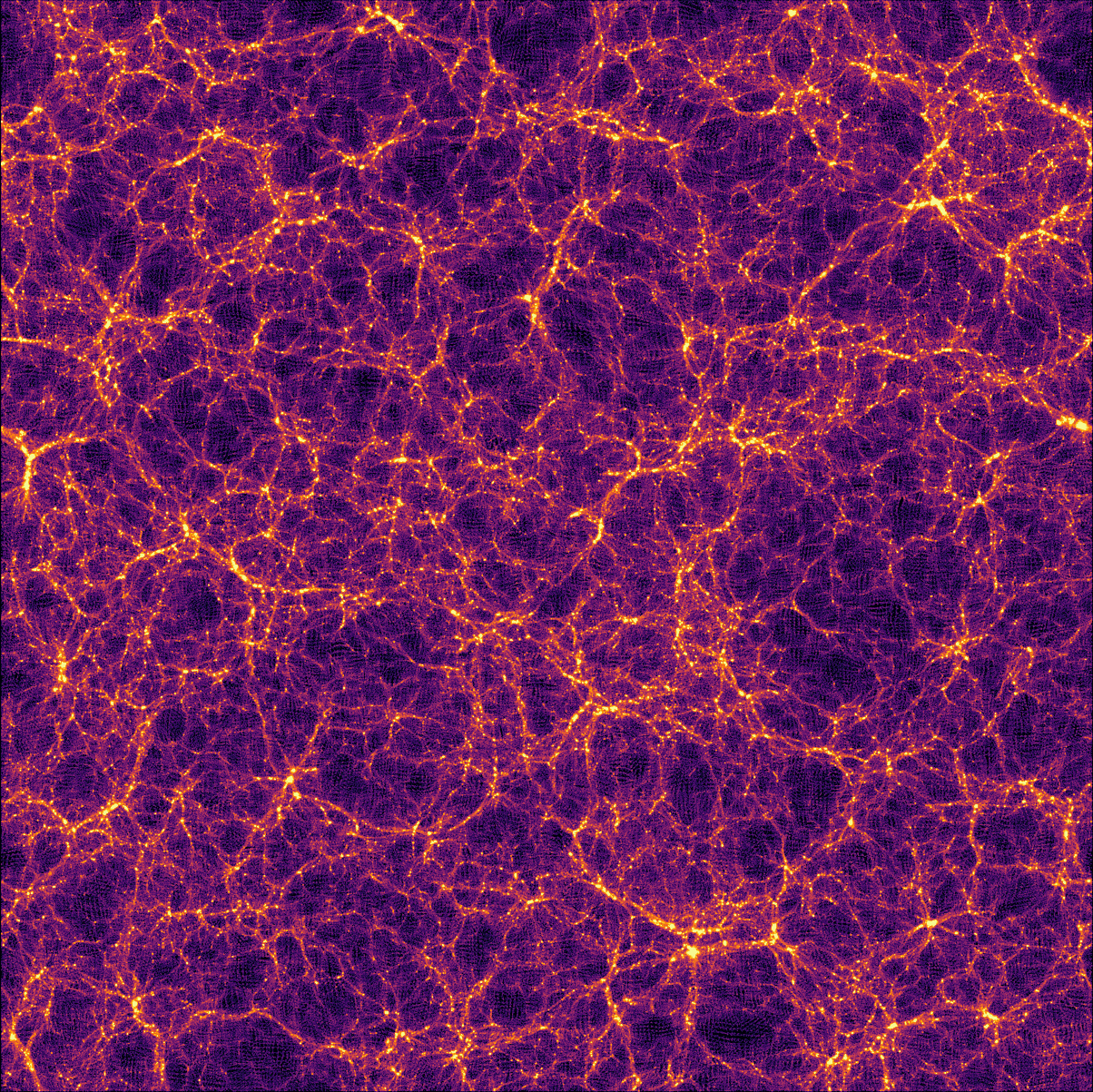}
      \hspace{10mm}
      \includegraphics[width=0.42\textwidth]{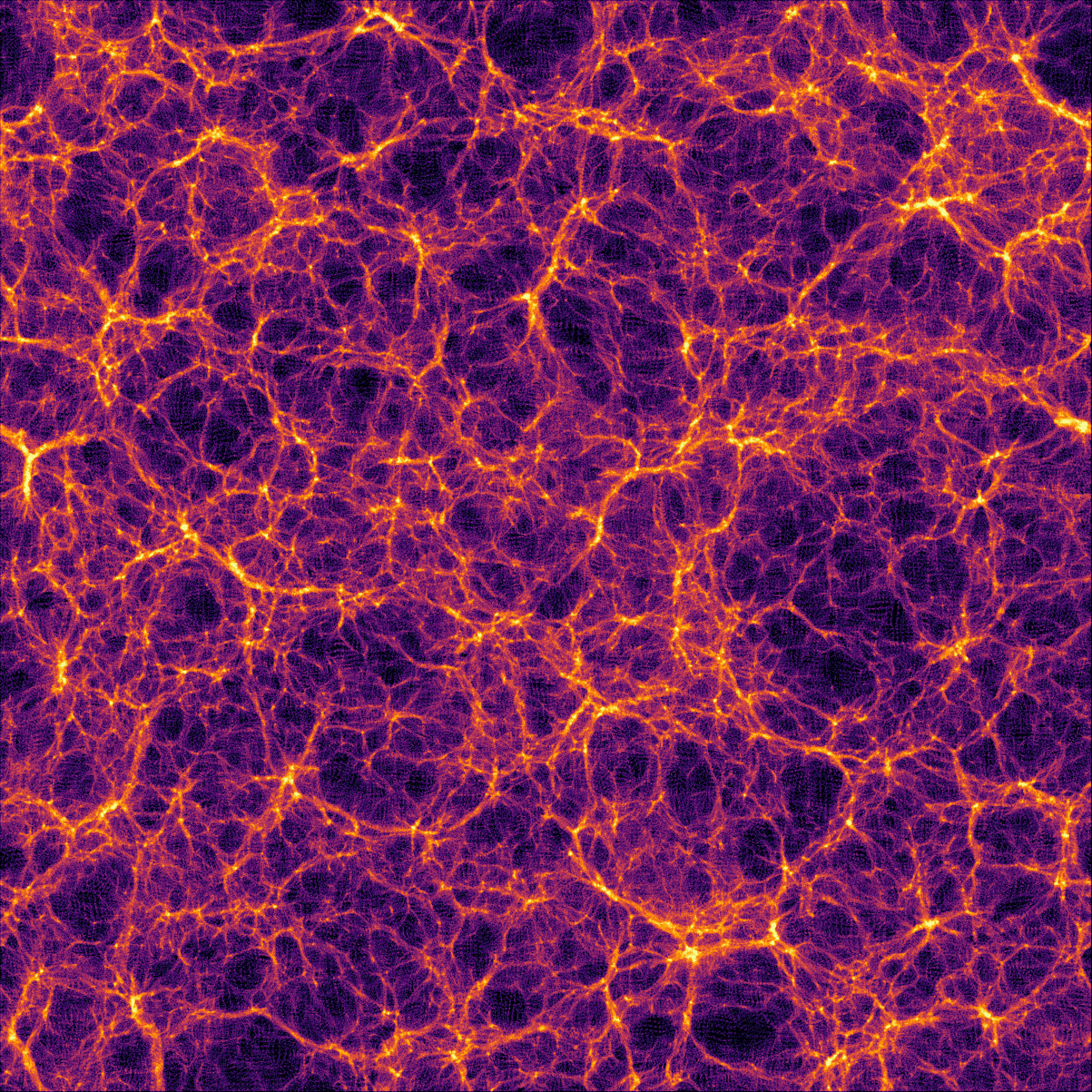}
    }
    \centerline{\color{white} $\Lambda$CDM, $z=3$ \hspace{55mm}Monge-Ampère, $z=3$}
    \centerline{
      \includegraphics[width=0.42\textwidth]{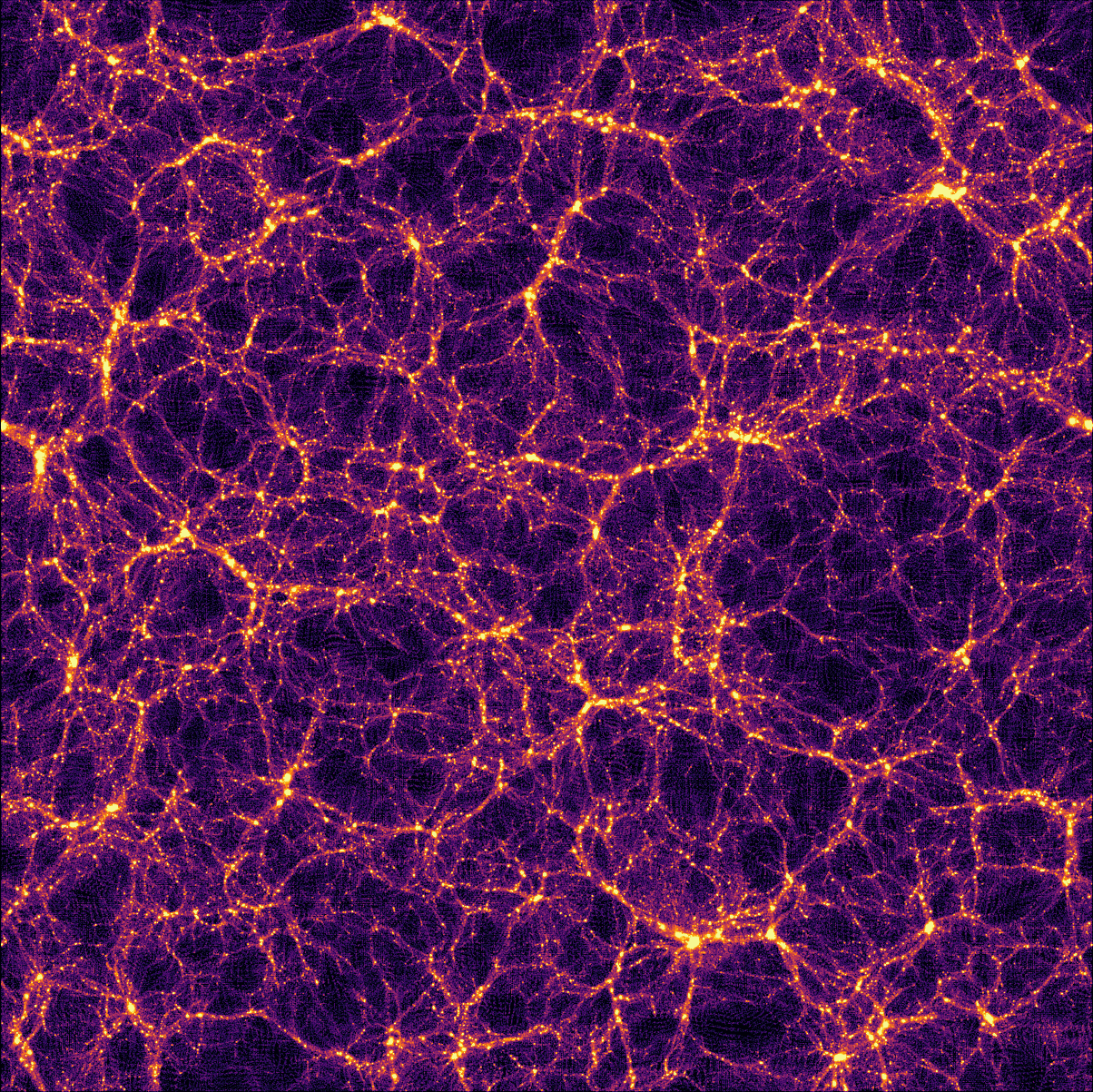}
      \hspace{10mm}
      \includegraphics[width=0.42\textwidth]{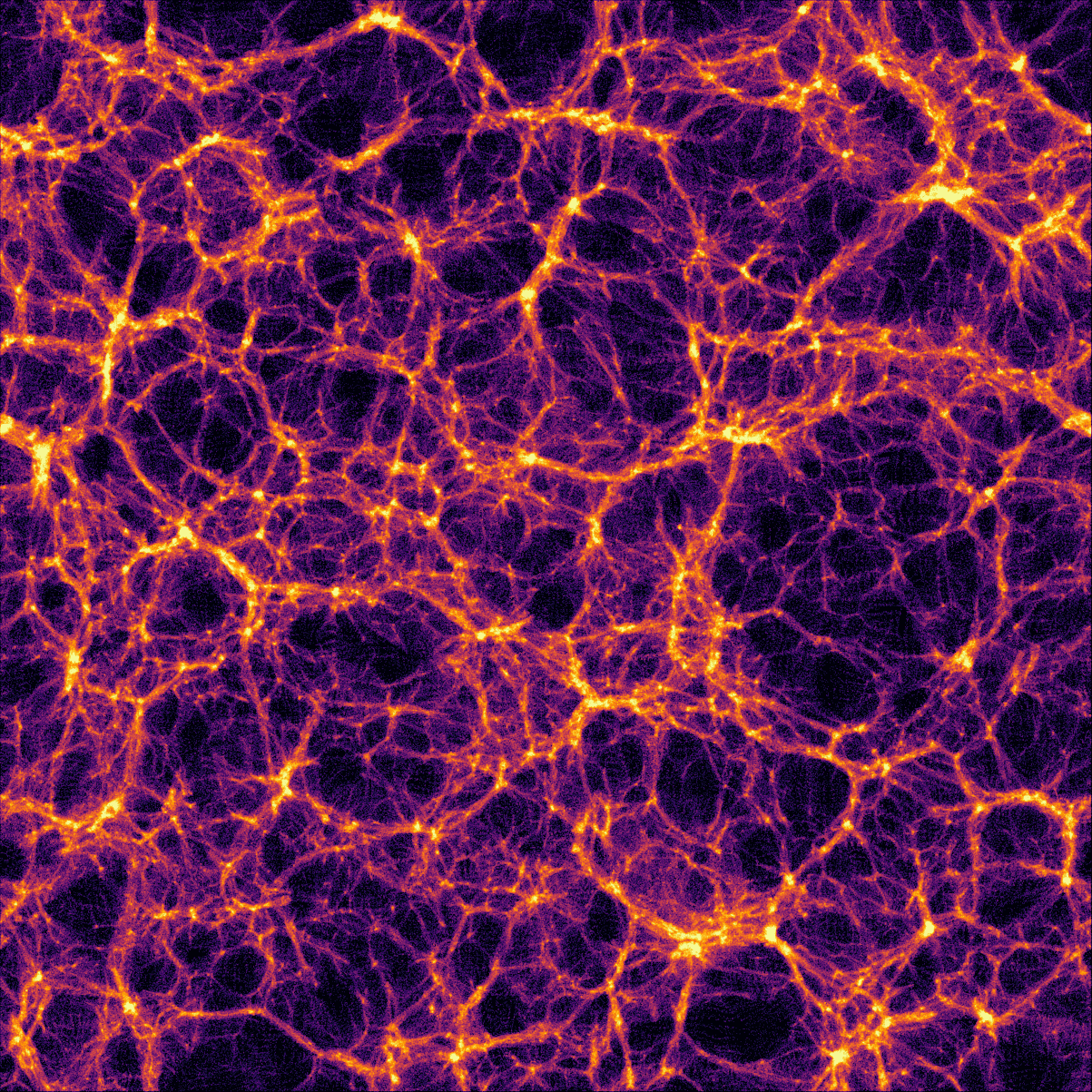}
    }
    \centerline{\color{white} $\Lambda$CDM, $z=0$ \hspace{55mm}Monge-Ampère, $z=0$}
    \end{minipage}
    }
    \caption{
    Comparison between \textcolor{black}{3D} simulations of $\Lambda$CDM (using an adaptive-mesh algorithm similar to \cite{Couchman1994HydraAA}) and Monge-Ampère in a
    cube of 300 Mpc/h, $512^3$ particles, z=5, 3 and 0.
    Projected integrated density in a 15 Mpc/h thick slab, using a logarithmic color scale. Large-scale similarity between the two models is striking, however
    MAG creates more abundant and diffuse filaments, whereas $\Lambda$CDM creates highly-clustered small haloes. There is weaker clustering because MAG does not diverge and is screened at short distances.}
    \label{fig:300Mpch}
\end{figure*}

\begin{figure}
    \centering
    \includegraphics[width=\columnwidth]{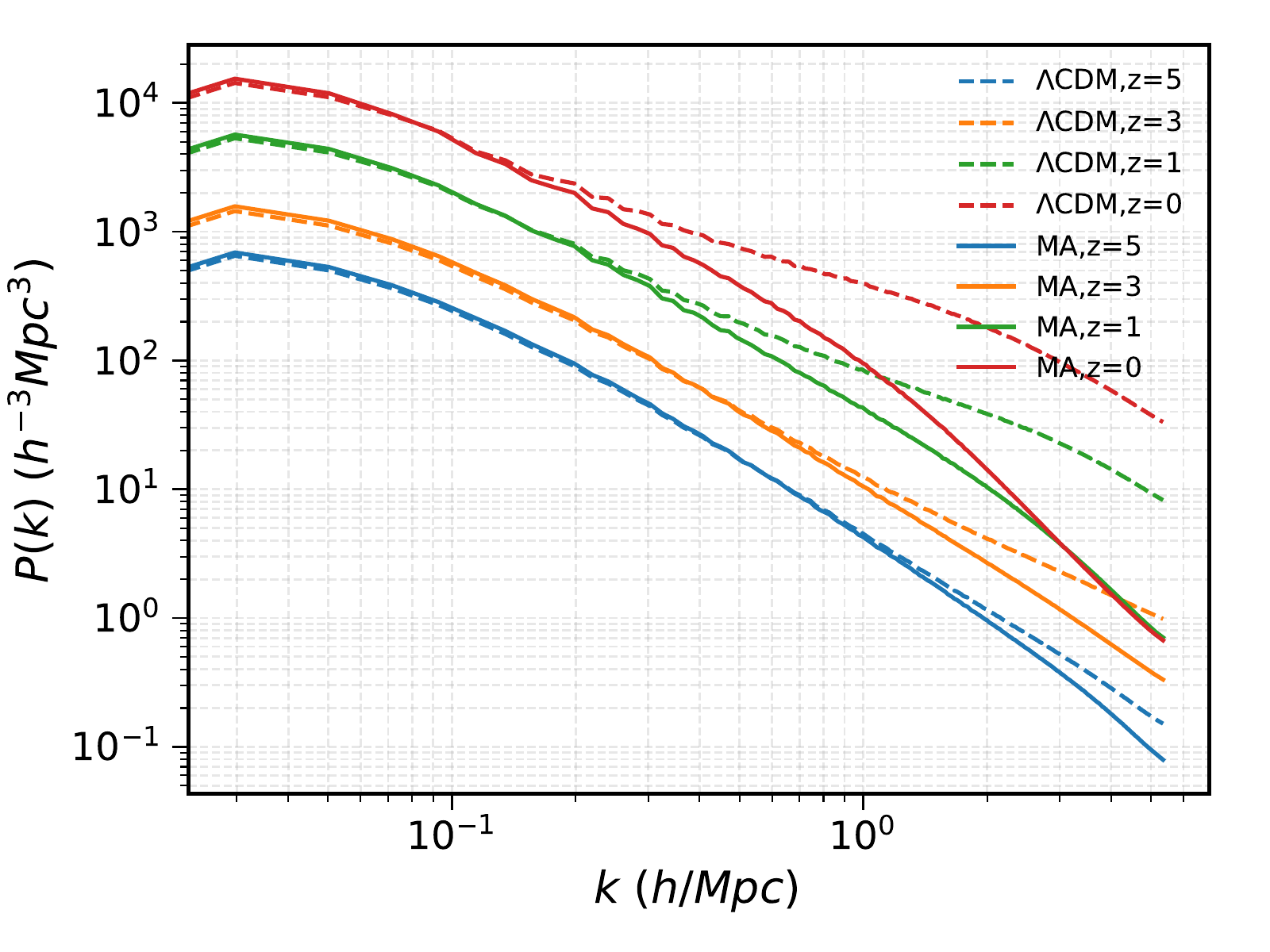}
    \caption{Power spectrum of $\Lambda$CDM and Monge-Ampère simulations (300 Mpc/h, $512^3$ particles).
    At large scales, $\Lambda$CDM and Monge-Ampère have the same trend (with slightly more power for Monge-Ampère). At small scales, Monge-Ampère is screened: it has less power than
    $\Lambda$CDM.}
    \label{fig:powerspectrum}
\end{figure}

\begin{figure*}

    \colorbox{black}{
    \begin{minipage}{\textwidth}

     \centerline{      \includegraphics[width=0.5\textwidth]{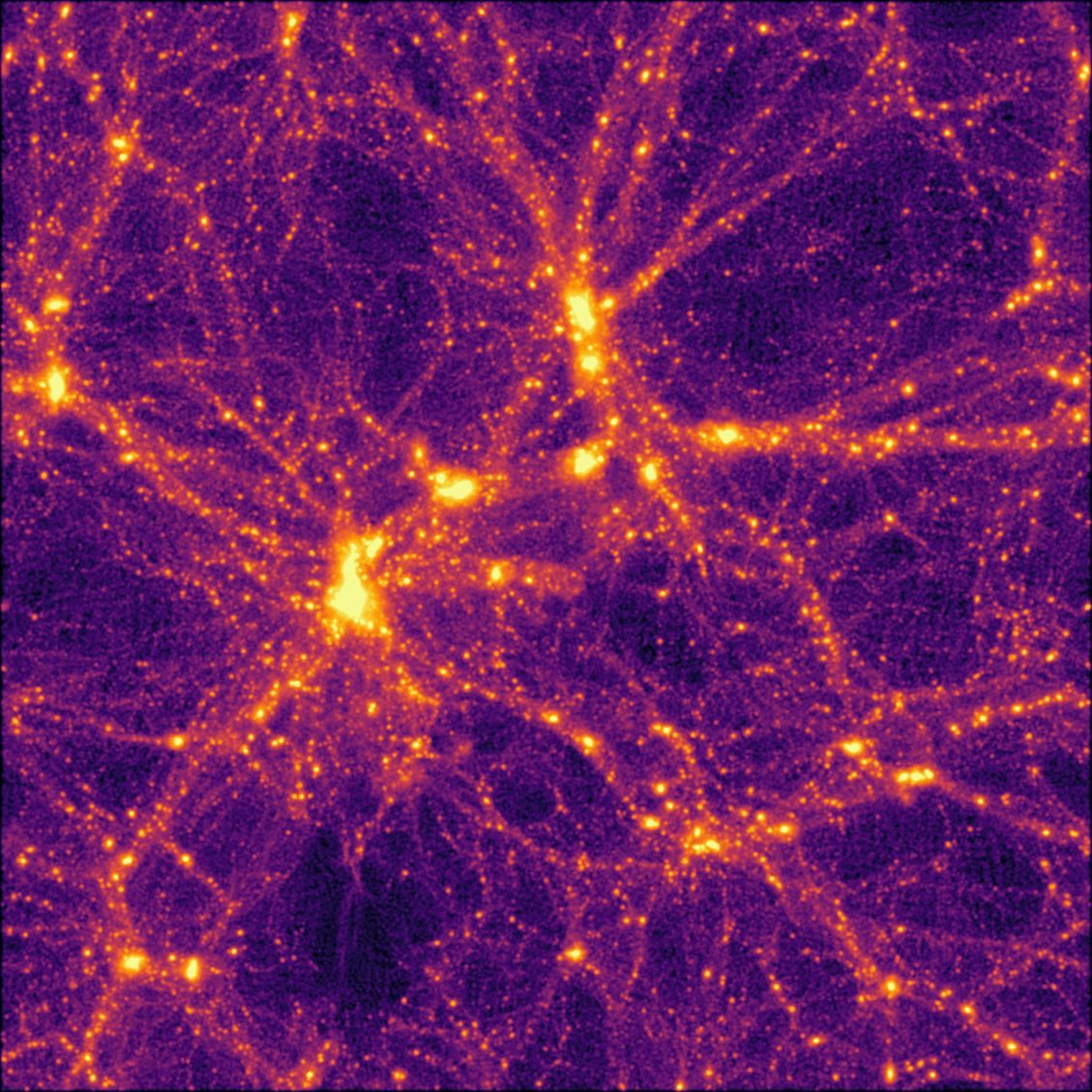}      \includegraphics[width=0.5\textwidth]{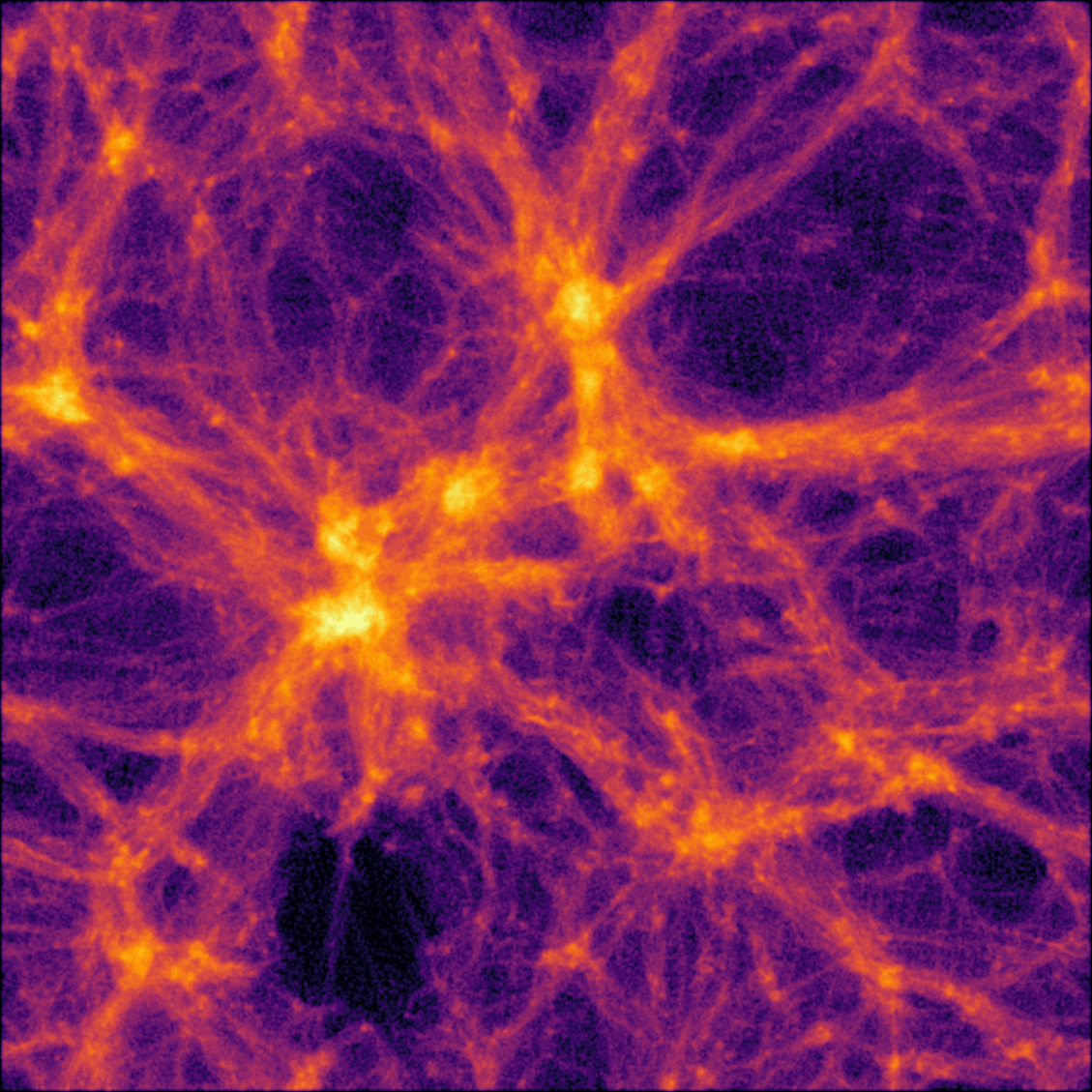}
    }
    \vspace{5mm}
    \centerline{      \includegraphics[width=0.5\textwidth]{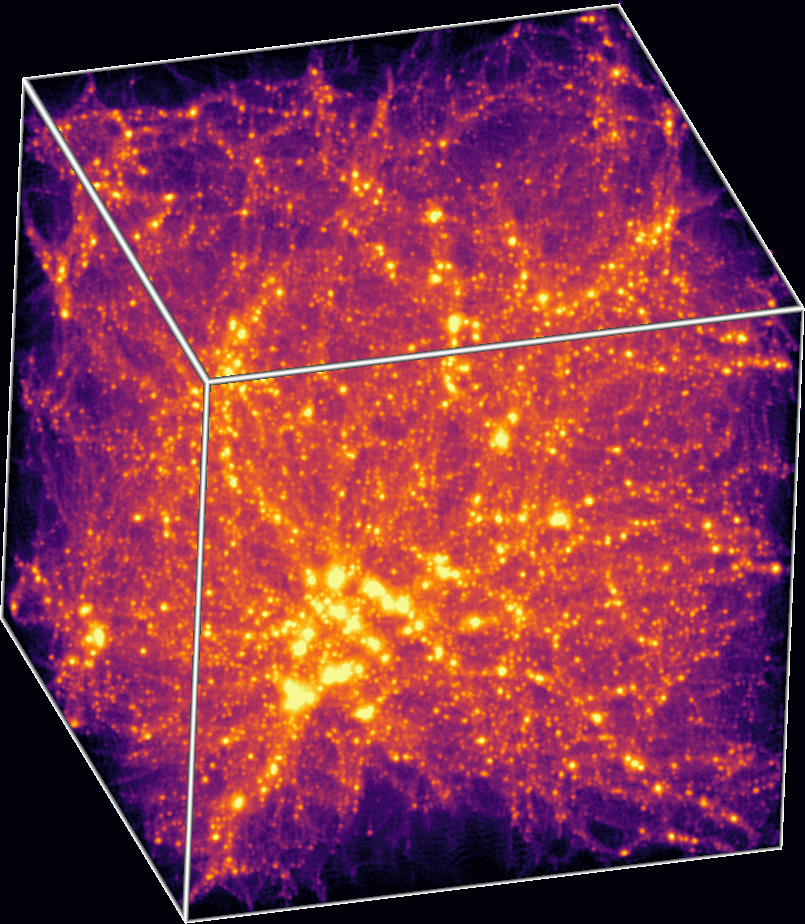}         \includegraphics[width=0.5\textwidth]{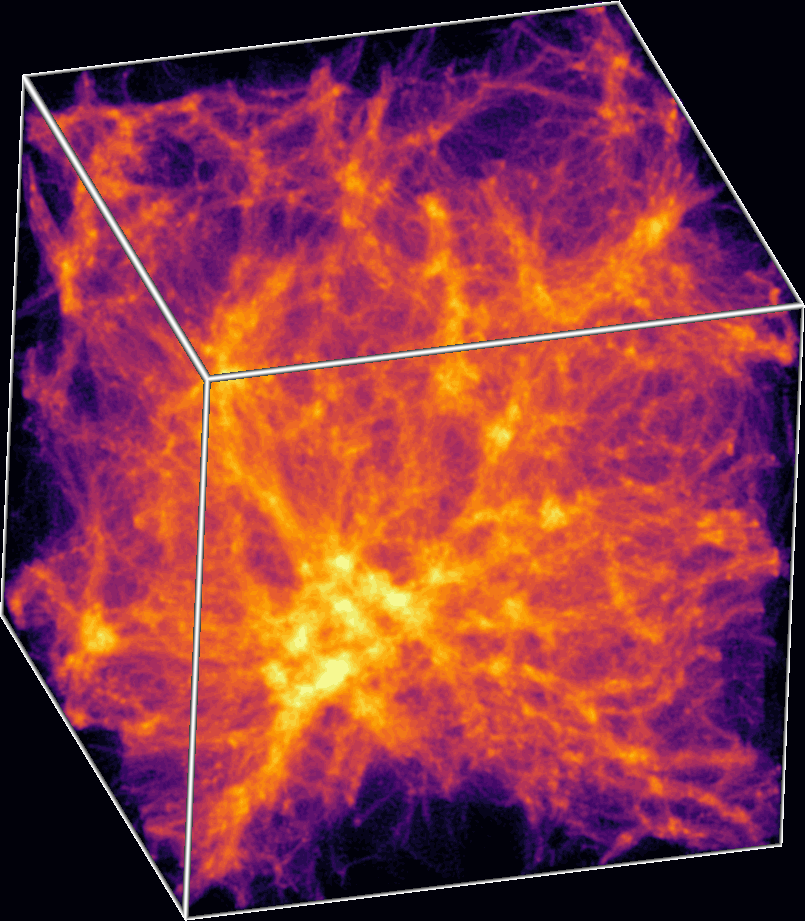}
    }

    \end{minipage}
    }

    \caption{
     Zoom-simulation in a cube of 60 Mpc/h, $256^3$ particles.
    Left column: $\Lambda$CDM simulation; Right column: Monge-Ampère simulation (z=0 for both). Top row shows the projected integrated density on a 1 Mpc/h thick slab. All images use a logarithmic color scale.
    In this ``zoom-simulation", one can better observe that Monge-Ampère gravity tends to create more diffuse and more anisotropic filaments, whereas $\Lambda$CDM creates a larger number of small dense haloes. Moreover, voids seem {\it emptier} in Monge-Ampère gravity. \textcolor{black}{(See also the companion paper \cite{albert} for more details and theoretical framework.)}
    }
    \label{fig:hydra16M}
\end{figure*}

\section{Simulation Results}
\label{sect:results}

We conducted numerical experiments to compare the effect of
Monge-Ampère gravity with Newtonian gravity in $\Lambda$CDM.
We ran two simulations: one with Newtonian dynamics and one with \MAG, both in comoving coordinates for an expanding Universe with periodic boundary conditions, with $512^3$ particles in a \textcolor{black}{3D} box of $300$ Mpc/h with $\Lambda$CDM initial condition and parameters from Planck collaboration \cite{planck2016}. The $\Lambda$CDM simulation uses an adpative-mesh algorithm similar to \cite{Couchman1994HydraAA}, and the Monge-Ampère simulation uses the algorithm described in the previous section \textcolor{black}{(see Figure \ref{fig:MA_3D})}.

Figure \ref{fig:300Mpch} shows a slab with a thickness of 15 Mpc/h displayed with a logarithmic color scale, for both $\Lambda$CDM and Monge-Ampère, at redshifts 5,3,1 and 0.

The large-scale structures have globally the same shape for both simulations, which was expected since Monge-Ampère can be considered as a small perturbation of Poisson. At a finer scale,
the Monge-Ampère simulation has more abundant and more diffuse filaments, especially at low redshift: in Newtonian gravity, filaments are clumpier and are transformed into multiple spherical regions of high density. The associated power spectra are shown in Figure \ref{fig:powerspectrum}.
 At small scales, Monge-Ampère has much less power than $\Lambda$CDM (that creates dense lumps of matter). At large scales, $\Lambda$CDM and Monge-Ampère have the same trend (with slightly more power for Monge-Ampère). This expected since in low density regions
far away from dense regions, the
nonlinearities are not important and we recover linear Poisson-like equation. On the other hand, in high-density regions near massive bodies, the nonlinear terms
become important and, a mechanism similar to the screening mechanism in modified theories of gravity ({\it e.g.}\ Vainshtein mechanism), effectively suppress the spatial derivative of the scalar field. Hence we observe fewer dense haloes in \MAG. (See also the companion paper \cite{albert}).

\begin{table}[]
    \centering
    \begin{tabular}{c|c|c|c|c|}
                      & Total time (s) & timesteps & time/timestep & Newton it. \\
                      \hline
         $\Lambda$CDM & $1.1047 \times 10^6$ & 1797 & 10 min 14 s & N/A \\
         MA   & $1.1343 \times 10^6 $ & 95 & 40 min to 7h20 & 2 to 12
    \end{tabular}
    \caption{Timings for the 300 Mpc/h simulation with $512^3$ particles, for the $\Lambda$CDM and the Monge-Ampère simulations.}
    \label{tab:timings}
\end{table}

In terms of computation time, as summarized in Table \ref{tab:timings}, our optimized multithreaded solver for Monge-Ampère took $1.1047\times10^6$ seconds (300 hours), on a Intel Xeon Gold 5122 at 3.6 GHz, with 100 Gb of RAM. Our $\Lambda$CDM simulation took approximately the same time, $1.1343\times10^6$ seconds (monothreaded code, similar to \cite{Couchman1994HydraAA}). Clearly, much faster results can be obtained for $\Lambda$CDM with recent multithreaded codes, such as GADGET-4 \cite{10.1093/mnras/stab1855} (at least 10x faster). It is expected that our algorithm will be significantly slower than modern $\Lambda$CDM codes, because to solve the Monge-Ampère equation, our Newton algorithm needs to solve \emph{multiple} discrete Poisson equations at each time step, between 2 to 12 in the present case (as we approach z=0, the density fields becomes more and more irregular, and the KMT Newton algorithm for semi-discrete optimal transport needs more and more iterations to converge). However, the very favorable CFL condition lets us compute much larger timesteps. This is made possible by both our unstructured representation based on Laguerre cells, and the fact that the solutions of Monge-Ampère equation are in general much more regular than Poisson solutions: densities become less clustered, and velocities remain bounded. The Monge-Ampère simulation reached $z=0$ after 94 timesteps, to be compared by the 1797 timesteps used by the $\Lambda$CDM simulation.
Clearly, this very favorable CFL coundition will not suffice to make our algorithm compete with fast $\Lambda$CDM codes (that need solving a \emph{single} Poisson equation per timestep). However, we wish to stress that computation times remain reasonable, and our algorithm is to our knowledge the first one that can be used in practice for high-resolution simulation of such highly non-linear modified theories of gravity. \\

To see the effect of Monge-Ampère gravity at a higher resolution, we also ran a ``zoom simulation in a cube of 60 Mpc/h, with $256^3$ particles (2.5x resolution in each direction in terms of number of particles). A slab with a thickness of 1 Mpc/h and a 3D view are both shown in Figure \ref{fig:hydra16M}. At this scale, one can better see the structure of the filaments, how they are {\it fragmented} by Newtonian gravity, and how they remain continuous while more fuzzy in Monge-Ampère gravity. In addition, one can see that the voids are emptier in Monge-Ampère gravity (this is also visible at $z=0$ in the 300 Mpc/h simulation in Figure \ref{fig:300Mpch}). \\

Animations of both the 300 Mpc/h and 60 Mpc/h simulations are available from \url{http://brunolevy.github.io/videos/index.html}

\section{Conclusions}

In this work, we have provided the first high-resolution numerical simulation of Monge-Ampère equation as an effective theory of gravity on large scales. Our algorithm based on semi-discrete optimal transport theory allows a novel high-resolution simulation of this highly nonlinear second-order differential equation. More quantitative numerical tests of the Monge-Ampère equation shall be presented in the forthcoming works, together with an extended algorithm for conducting higher-resolution numerical experiments.

We shall conclude by discussing the new lines of research that follow our results, more precisely, we ask the following question:

{\it What is the physical significance of Monge-Ampère equation and does it have any bearing on modified theories of gravity ?}
We outline three possible scenarios in which Monge-Ampère equation can describe gravity.

In the {\it first scenario}, which we have explored in this work, MAG is not a fundamental but an effective theory of gravity on cosmological scales.
The solar system provides one of the most stringent tests of theory of general relativity. It is clear that MAG cannot replace Poisson equation as a fundamental force, as the nonlinearities of the Monge-Ampère equation would violate the solar system test \cite{bonnefous}. MAG yields Poisson in the weak gravity regime but will not yield Poisson in strong gravity regime, {\it e.g.} in the solar system. Here, we have shown that MAG can {\it emerge} from a system of Brownian indistinguishable and independent particles through the application of large-deviation principle. We have ran parallel simulations with Poisson equation and Monge-Ampère equation. Comparison between these two simulations show that MAG supresses small scale fluctuations, and favours the formation of anisotropic structures such as filaments. We understand this
to be due to nonlinear terms that supress clustering at small scales where MAG does not diverge. Furthermore, the Monge-Ampère equation has a higher symmetry group, invariant through a shearing (GL(3)), whereas the
Poisson equation only has rotational invariance (SO(3)). The affine invariance of the Monge-Ampère equation implies that the whole range of new solutions obtained through shearing and deformation are stable.  In addition, the Laplacian, tends to homogenise the fluctuations and shift power from large to small scales. We emphasise that, Monge-Ampère gravity is weaker at small scales and hence yields less clustering (compare to \cite{li} and see also the companion paper \cite{albert}).

In {\it the second scenario}, MAG is a fundamental theory of gravity that describes a scalar field which is coupled to matter with the latter satisfying the Einstein theory of general relativity, {\it i.e.}\ a scalar-tensor theory.
In this scenario the Monge-Ampère equation represents the equation of motion of a large class of Lagrangians, such as the Galileons and those related to extra-dimensional theories, and can be considered as the fifth force (see {\it e.g.} \cite{fairlie-comments,fairlie-premier,gilles,khoury,nicolis,burrage,deffayet-steer,deser}). The scalar field is naturally screened at short distances (similar to Vainshtein mechanism), {\it i.e.}\ at the solar system scale, because of nonlinearies and at very large distances the Newtonian gravity prevails. The effect of scalar-field is most prominent at galactic scales. \textcolor{black}{The physics behind Monge-Ampère equation as a model for scalar fields is discussed in a  companion article \cite{bonnefous}. In this work it was shown that the Monge-Ampère equation is the field equation of quartic Galileon with different interesting properties \cite{gilles}. It was also show that the equation of Monge-Ampère gravity, which uses \eqref{eqn:potMA} and  \eqref{eqn:potMA2}, arises as the equation of motion of the sum of all Galileons. The reader is referred to the companion paper for full details \cite{albert}.}

In {\it the third scenario}, MAG is a fundamental theory of dark matter whereas the Poisson equation describes the luminous matter. Models in which the equivalence principle between dark and luminous matter is broken, have been studied before in different contexts (see {\it e.g.} \cite{uzan}) and is discussed in the companion work \cite{bonnefous}.

Finally, the major motivation for introducing a scalar field into theories of gravity is to produce an alternative to dark energy and to remedy the cosmological constant problem. Here, we have ran our simulations in a FLRW background. The study of growth rate in MAG as an alternative fundamental theory of gravity will be carried in the forthcoming work.


\section*{acknowledgements}

Bruno Lévy's work was supported by the Inria Exploratory Project Grant COSMOGRAM-launchpad.

We mention the existence of other statistical measures of our simulated Monge-Ampère gravity done by P. Boldrini in the frame of his Post-Doc and then in collaboration with C. Laigle, that they published in \cite{boldrini2024distinguishdarkmattertheories}.

We wish to thank Mike Cullen, Gilles Esposito-Faerese, Cédric Deffayet, Christian Léonard, André Lukas,  Quentin Mérigot, Farnik Nikakhtar, Henri Orland, Pierre Salati,  Jean-Philippe Uzan and Sebastian von Hausegger for discussions, and Denis Demidov for his help with AMGCL.

\bibliography{BMAG}
\end{document}